\documentclass[a4paper,12pt]{article}

\usepackage{graphicx}
\usepackage{array}
\usepackage{amsmath}
\usepackage{amssymb}
\usepackage{latexsym}
\usepackage{subfigure}
\usepackage{color}

 \makeatletter
 \@addtoreset{equation}{section}
 \makeatother

\date{empty}
\begin{document}
\begin{titlepage}
\null
\begin{flushright}
April, 2015
\end{flushright}
\vskip 1.5cm
\begin{center}
  {\Large \bf Classifying BPS States in 
Supersymmetric Gauge Theories
Coupled to Higher Derivative Chiral Models 
}
\vskip 1.5cm
\normalsize
\renewcommand\thefootnote{\alph{footnote}}

{\large
Muneto Nitta$^{\dagger}$\footnote{nitta(at)phys-h.keio.ac.jp}
and Shin Sasaki$^\ddagger$\footnote{shin-s(at)kitasato-u.ac.jp}
}
\vskip 0.7cm
  {\it
  $^\dagger$ 
Department of Physics, and Research and Education Center for Natural Sciences, \\
\vskip -0.2cm
Keio University, Hiyoshi 4-1-1, Yokohama, Kanagawa 223-8521, Japan  
\vskip 0.1cm
$^\ddagger$
  Department of Physics,  Kitasato University \\
  \vskip -0.2cm
  Sagamihara 252-0373, Japan
}
\vskip 0.5cm
\begin{abstract}
We study $\mathcal{N} = 1$ supersymmetric gauge theories  coupled with 
higher derivative chiral models 
in four dimensions 
in the off-shell superfield formalism. 
We solve the equation of motion for the auxiliary fields and find 
two distinct on-shell structures of the Lagrangian 
that we call the canonical and non-canonical branches
characterized by zero and non-zero auxiliary fields, respectively. 
We classify BPS states of the models in Minkowski and Euclidean spaces.
In Minkowski space, we find Abelian and non-Abelian vortices, 
vortex-lumps (or gauged lumps with fractional lump charges)
as 1/2 BPS states in the canonical branch and 
higher derivative generalization of vortices and 
vortex-(BPS)baby Skyrmions 
(or gauged BPS baby Skyrmions with fractional baby Skyrme charges)
as 1/4 BPS states in the non-canonical branch.
In four-dimensional Euclidean space, we find 
Yang-Mills instantons trapped inside a non-Abelian vortex,  
intersecting vortices, and intersecting 
 vortex-(BPS)baby Skyrmions as 1/4 BPS states 
in the canonical branch but no BPS states in the non-canonical branch 
other than those in the Minkowski space.

\end{abstract}
\end{center}

\end{titlepage}

\newpage
\setcounter{footnote}{0}
\renewcommand\thefootnote{\arabic{footnote}}
\pagenumbering{arabic}

\newpage
\section{Introduction}

Low-energy effective theories play an important role 
in the study of non-perturbative effects of quantum field theory, 
such as the chiral Lagrangian of QCD \cite{Leutwyler:1993iq}. 
In certain supersymmetric gauge theories, low-energy effective theories are 
determined exactly offering full quantum spectra of 
Bogomol'nyi-Prasado-Sommerfield (BPS) states 
\cite{Seiberg:1994rs}. 
BPS states 
preserve a part of supersymmetry, 
belonging to so-called short multiplets 
of supersymmetry algebra,
and consequently they are stable against quantum corrections 
perturbatively and non-perturbatively 
\cite{Witten:1978mh}.  
The low-energy effective field theories 
are constructed by a derivative expansion 
and are usually complemented by  
higher derivative corrections, 
as in the chiral perturbation theory  
\cite{Leutwyler:1993iq}.

Recently, in our previous paper, 
BPS states in the supersymmetric chiral models with 
higher derivative terms have been classified 
in ${\cal N}=1$ supersymmetric theories in four dimensions 
\cite{Nitta:2014pwa}. 
The purpose of this paper 
is to classify BPS states in 
 ${\cal N}=1$  supersymmetric gauge theories 
coupled with higher derivative chiral models 
in four-dimensional Minkowski and Euclidean spaces.
  
Higher derivative corrections to supersymmetric field theories have 
a long history because of the auxiliary field problem.
The auxiliary fields $F$ in the off-shell superfield formalism of higher derivative models
are generically acted on by space-time derivatives 
and consequently cannot be eliminated algebraically 
to obtain on-shell actions. 
Supersymmetric higher derivative terms 
free from the auxiliary field problem 
have been studied individually in various contexts:
the Wess-Zumino-Witten term  
\cite{Nemeschansky:1984cd,Gates:1995fx,Gates:2000rp,Nitta:2001rh},
low-energy effective action 
\cite{Buchbinder:1994iw, Buchbinder:1994xq, Matone:1996bj, Bellisai:1997ck,
Gomes:2009ev, Banin:2006db, AnDuGh, Gama:2011ws, Kuzenko:2014ypa},
${\mathbb C}P^1$ (Faddeev-Skyrme) model
\cite{BeNeSc,Fr}, 
Dirac-Born-Infeld (DBI) action 
\cite{RoTs,SaYaYo},
$k$-field theory
\cite{AdQuSaGuWe3,AdQuSaGuWe2},
low-energy effective action on BPS solitons \cite{Eto:2012qda}, 
BPS baby Skyrme model 
\cite{Adam:2011hj,AdQuSaGuWe,Nitta:2014pwa,Bolognesi:2014ova}, 
and nonlinear realizations of 
Nambu-Goldstone fields \cite{Nitta:2014fca}.
 In the framework of supergravity,  
higher derivative terms
\cite{KhLeOv,KoLeOv2,KhLeOv2,KoLeOv,FaKe} 
have been applied to 
ghost condensations \cite{KhLeOv,KoLeOv2} 
and the Galileon inflation models \cite{KhLeOv2}. 
Among those, the four derivative term first found in Ref.~\cite{Buchbinder:1994iw},  that can be constructed from 
a $(2,2)$ K\"ahler tensor, 
was rediscovered in Refs.~\cite{KhLeOv,KoLeOv2} 
and has recently been used in various contexts. 
By using a K\"ahler tensor containing 
space-time derivatives, 
one can construct   higher derivative terms with 
arbitrary number of space-time derivatives 
\cite{Nitta:2014fca}.

In our previous paper \cite{Nitta:2014pwa},
the auxiliary field equations 
were found to admit at least two distinct solutions 
that we called 
canonical and non-canonical branches 
with $F=0$ and $F\neq 0$, respectively. 
In particular, 
BPS baby Skyrmions (compactons) 
 \cite{Adam:2011hj,AdQuSaGuWe} 
have been found to be 1/4 BPS states 
in the non-canonical branch,
while BPS lumps  are 1/2 BPS states 
in the canonical branch \cite{Eto:2012qda},
although both of them saturate the same Bogomol'nyi bound. 
In the former, the on-shell Lagrangian contains  
no usual kinetic term and consists of only a four derivative term,  
while in the latter, higher derivative corrections disappear 
in solutions and energy. 
BPS baby Skyrmions as compactons are currently 
paid much attention \cite{Adam:2009px,Adam:2012pm}.

In this paper, 
we classify BPS states in ${\cal N}=1$  supersymmetric gauge theories 
coupled with higher derivative chiral models 
in four-dimensional Minkowski and Euclidean spaces.
Here, we concentrate on the cases where superpotentials are absent for simplicity. 
As in the previous cases without gauge fields, 
we find canonical and non-canonical branches
corresponding to solutions $F=0$ and $F\neq 0$ 
of auxiliary field equations, respectively. 
We find that 
1/2 BPS states that exist 
in theories without higher derivative terms 
remain 1/2 BPS  in the canonical branch  
and that corresponding BPS states in the non-canonical branch 
are 1/4 BPS states.
On the other hand, we also find that 
1/4 BPS states that exist 
in theories without higher derivative terms 
remain 1/4 BPS in the canonical branch 
but there are no corresponding BPS states in the non-canonical branch.
More precisely, 
we find that 1/2 BPS equations in the canonical branch 
do not receive higher derivative corrections 
for an Abrikosov-Nielsen-Olesen (ANO) vortex 
\cite{Abrikosov:1956sx} 
at the critical (BPS) coupling, 
a non-Abelian vortex \cite{Hanany:2003hp},  
lumps \cite{Polyakov:1975yp}, 
vortex-lumps 
(gauged lumps with fractional lump charges) \cite{Schroers:1995he,Nitta:2011um}. 
We then show that higher derivative generalization of vortices 
(that we may call compact vortices) 
and vortex-baby Skyrmions 
(or gauged baby Skyrmions with fractional baby Skyrme charges)
are 1/4 BPS states   
in the non-canonical branch. 
In  four-dimensional Euclidean space, we find 1/2 BPS Yang-Mills instantons,
1/4 BPS Yang-Mills instantons trapped inside a non-Abelian vortex, 
and 1/4 BPS intersecting vortices with instanton charges in the canonical branch.
These configurations were known in supersymmetric theories with 
eight supercharges without higher derivative terms 
in $4+1$ or $5+1$ dimensions \cite{Hanany:2004ea,Eto:2004rz,Eto:2006pg,Fujimori:2008ee}, 
and so what we confirm here is that they are still 1/4 BPS states 
in theories with four supercharges
in Euclidean four dimensions and that 
higher derivative terms are canceled out 
in the BPS equations and energy bound.
Further, as new configurations, we find 1/4 BPS vortex-lump string intersections with Yang-Mills instanton charges.
We find no BPS states in the non-canonical branch other than those in Minkowski space. 

This paper is organized as follows. 
In Sec.~\ref{sec:hdcm}, we give supersymmetric Lagrangian 
in the superfield formalism.
The first subsection is devoted to a review for 
higher derivative chiral models of chiral multiplets 
without coupling to gauge fields.
In the second subsection, we introduce vector multiplets 
and coupling of vector and chiral multiplets.
In Sec.~\ref{sec:BPS-Minkowski} we classify BPS states 
in four-dimensional Minkowski space.
In Sec.~\ref{sec:BPS-Euclidean}
BPS states in four-dimensional Euclidean space are discussed.
Sec.~\ref{sec:summary} is devoted to a summary and discussion.
Notations and conventions are summarized in Appendix.~\ref{sec:notation}.
Explicit supersymmetry variations of fermions in Euclidean space are 
found in Appendix.~\ref{sec:SUSY_variation}

\section{Higher derivative chiral model}\label{sec:hdcm}
In this section we introduce the four-dimensional $\mathcal{N} = 1$ supersymmetric higher
derivative chiral model \cite{KhLeOv, Nitta:2014pwa} and its coupling to
the vector multiplet. 
The supersymmetric higher derivative chiral model consists of chiral
superfields $\Phi^i \ (i = 1, \ldots, n)$ with an arbitrary K\"ahler potential $K$, 
superpotential $W$ and a symmetric $(2,2)$ K\"ahler tensor
$\Lambda_{ik\bar{j} \bar{l}}$. The tensor $\Lambda_{ik\bar{j}
\bar{l}}$ is an arbitrary function of $\Phi^i, \Phi^{\dagger \bar{j}}$
and its space-time derivatives.
Among other things, the purely bosonic part of the model never contains
the space-time derivatives of the auxiliary fields $F^i$.
Then all the auxiliary fields are integrated out by the algebraic equation of
motion and one finds explicit on-shell Lagrangians.
When global symmetries in the model are gauged, the higher derivative term couples to
the vector multiplet. 
In the following, we provide the explicit Lagrangian of the non-gauged higher
derivative chiral model and its coupling to the vector multiplet (gauged
model).

\subsection{Higher Derivative Chiral Models without Gauge Coupling}
We first start from the non-gauged $\mathcal{N} = 1$ supersymmetric 
higher derivative model with chiral superfields $\Phi^i$.
We employ the Wess-Bagger convention \cite{Wess:1992cp} in this paper
and detailed conventions and notations are summarized in Appendix \ref{sec:notation}.
The component expansion of the chiral superfield in the chiral base 
$y^m = x^m + i \theta \sigma^m \bar{\theta}$ is 
\begin{align}
\Phi^i = \varphi^i (y) + \theta \psi^i (y) + \theta^2 F^i (y).
\end{align}
Here $\varphi^i$ is the complex scalar field, $\psi^i$ is the Weyl
fermion, and $F^i$ is the  auxiliary complex scalar field.
The Lagrangian of the non-gauged higher derivative chiral model is given by 
\begin{align}
\mathcal{L} = & \
\ \int \! d^4 \theta \ K (\Phi^i, \Phi^{\dagger \bar{j}}) 
+ \frac{1}{16} \int \! d^4 \theta \ \Lambda_{i \bar{j} k \bar{l}} (\Phi,
 \Phi^{\dagger}) 
D^{\alpha} \Phi^i
 D_{\alpha} \Phi^k \bar{D}_{\dot{\alpha}} \Phi^{\dagger \bar{j}}
 \bar{D}^{\dot{\alpha}} \Phi^{\dagger \bar{l}} 
 + \left(\int \! d^2 \theta \ W(\Phi^i) + ({\rm h.c.})\right)
\label{eq:ungauged_HD_Lagrangian}
\end{align}
where $K$ is the K\"ahler potential, $\Lambda_{ik\bar{j}\bar{l}}$ is a
symmetric $(2,2)$ K\"ahler tensor and $W$ is the superpotential.
The fourth derivative part in the Lagrangian is evaluated as 
\begin{align}
D^{\alpha} \Phi^i D_{\alpha} \Phi^k \bar{D}_{\dot{\alpha}}
 \Phi^{\dagger \bar{j}} \bar{D}^{\dot{\alpha}}
 \Phi^{\dagger \bar{l}} 
=& \ 16 \theta^2 \bar{\theta}^2 
\left[
\frac{}{}
(\partial_m \varphi^i \partial^m \varphi^k) (\partial_n
 \bar{\varphi}^{\bar{j}} \partial^n \bar{\varphi}^{\bar{l}})
\right. 
\notag \\
& 
\left.
- \frac{1}{2} 
\left(
\partial_m \varphi^i F^k + F^i \partial_m \varphi^k 
\right)
\left(
\partial^m \bar{\varphi}^{\bar{j}} \bar{F}^{\bar{l}} 
+ \bar{F}^{\bar{j}} \partial^m \bar{\varphi}^{\bar{l}}
\right)
+ F^i \bar{F}^{\bar{j}} F^k \bar{F}^{\bar{l}}
\right] + I_f.
\label{eq:fourth_deri}
\end{align}
Here $I_f$ stands for terms that contain fermions.
Since the purely bosonic part in Eq.~\eqref{eq:fourth_deri} saturates the
Grassmann coordinate, only the lowest components in $\Lambda_{ik \bar{j}
\bar{l}}$ contribute to the bosonic part of the Lagrangian.
Then, the bosonic part of the Lagrangian is 
\begin{eqnarray}
\mathcal{L}_b &=& 
g_{i \bar{j}}
\left(
- \partial_m \varphi^i \partial^m \bar{\varphi}^{\bar{j}} + F^i \bar{F}^{\bar{j}}
\right)
+ \frac{\partial W}{\partial \varphi^i} F^i + \frac{\partial
\bar{W}}{\partial \bar{\varphi}^{\bar{j}}} \bar{F}^{\bar{j}}
\nonumber \\
& & + 
\Lambda_{ik\bar{j}\bar{l}} (\varphi, \bar{\varphi})
\left\{
(\partial_m \varphi^i \partial^m \varphi^k) (\partial_n \bar{\varphi}^{\bar{j}}
\partial^n \bar{\varphi}^{\bar{l}}) - 2 \partial_m \varphi^i F^k \partial^m
\bar{\varphi}^{\bar{j}} \bar{F}^{\bar{l}} + F^i \bar{F}^{\bar{j}} F^k \bar{F}^{\bar{l}}
\right\}.
\label{eq:off-shell_Lagrangian}
\end{eqnarray}
Here $g_{i\bar{j}} = \frac{\partial^2 K}{\partial \varphi^i \partial
\bar{\varphi}^{\bar{j}}} > 0$ is the K\"ahler metric.
In order to find the on-shell Lagrangian, we integrate out the auxiliary
fields $F^i$.
Since the Lagrangian does not contain space-time derivatives of the auxiliary
fields $F^i$, one can solve the equation of motion for $F^i$ and find the
explicit form of the purely bosonic part of the on-shell Lagrangian\footnote{
There are space-time derivatives of the auxiliary fields $F^i$ in the
fermion term $I_f$. Solutions to $F^i$ that include fermions are
obtained order by order of the fermions. 
Since we are interested in the classical configurations of fields, 
these fermionic contributions are irrelevant in this paper.
}.
The equation of motion for the auxiliary fields is 
\begin{align}
g_{i \bar{j}}
 F^i 
- 2 \partial_m \varphi^i F^k \Lambda_{ik\bar{j}\bar{l}} \partial^m
 \bar{\varphi}^{\bar{l}} + 2 \Lambda_{ik\bar{j}\bar{l}} F^i F^k
 \bar{F}^{\bar{l}} + \frac{\partial \bar{W}}{\partial
 \bar{\varphi}^{\bar{j}}} = 0.
\label{eq:auxiliary_eom}
\end{align}
As we have advertised, the equation \eqref{eq:auxiliary_eom} is an algebraic equation and it can be
solved in principle. There are distinct on-shell branches associated with different
solutions to the equation \eqref{eq:auxiliary_eom}.
In general, there are two classes of solutions. 
The first class has smooth limit $\Lambda_{ik\bar{j}\bar{l}} \to 0$ to the ordinary ({\it i.e.}~without higher
derivative terms) theory. For this class of solutions, higher derivative terms are
introduced as perturbations to the ordinary (with second space-time
derivatives) theory in the on-shell Lagrangian.
We call this case the canonical (perturbative) branch. 
On the other hand, the second class of solutions does not have 
a smooth limit $\Lambda_{ik\bar{j} \bar{l}} \to 0$ 
to the ordinary theory. For this class of
solutions, the higher derivative terms enter into the on-shell
Lagrangian non-perturbatively. 
We call this case the non-canonical (non-perturbative) branch.
In Ref.~\cite{Nitta:2014pwa}, we studied on-shell structures of the
Lagrangian \eqref{eq:off-shell_Lagrangian} for the single chiral superfield
model. When $W \not= 0$, the equation of motion for the auxiliary field
becomes that of the cubic power of $F$, 
and the solutions can be obtained by Cardano's method \cite{SaYaYo}.
The explicit solutions are quite non-linear in 
$K$, $\Lambda$, $W$, and $\partial_m \varphi$. Therefore, the on-shell Lagrangian
becomes a highly complicated function of the scalar field $\varphi$.
In the following, we consider models with $W = 0$ and show the explicit 
on-shell Lagrangians in the canonical and non-canonical branches.

\paragraph{Canonical branch}
It is apparent that $F^i = 0$ is always a solution to the equation 
\eqref{eq:auxiliary_eom}.
In this case, the bosonic part of the on-shell Lagrangian is 
\begin{align}
\mathcal{L}_b = - g_{i \bar{j}} \partial_m \varphi^i \partial^m
 \bar{\varphi}^{\bar{j}}
+ \Lambda_{ik \bar{j}\bar{l}} (\varphi, \bar{\varphi}) 
(\partial_m \varphi^i \partial^m \varphi^k) (\partial_n
 \bar{\varphi}^{\bar{j}} \partial^n \bar{\varphi}^{\bar{l}}).
\end{align}
The tensor $\Lambda_{ik \bar{j} \bar{l}}$ determines higher derivative
terms in the Lagrangian. 
Since $\Lambda_{ik \bar{j} \bar{l}}$ is an arbitrary function
of $\varphi, \bar{\varphi}$, one can construct arbitrary higher
derivative terms for $n=1$ models.
For example, the scalar part of the $\mathcal{N} = 1$ 
supersymmetric Dirac-Born-Infeld action \cite{RoTs} is obtained by the
single chiral superfield model with a flat K\"ahler potential and 
\begin{align}
\Lambda = \frac{1}{1 + A + \sqrt{(1 + A^2) - B}}, \quad 
A = \partial_m \Phi \partial^m \Phi^{\dagger}, \quad 
B = \partial_m \Phi \partial^m \Phi \partial_n \Phi^{\dagger} \partial^n
 \Phi^{\dagger}.
\label{eq:Lambda_DBI}
\end{align}
The supersymmetric Faddeev-Skyrme model is obtained by the
$\mathbb{C}P^1$ Fubuni-Study metric $K_{\varphi \bar{\varphi}} = \frac{1}{(1 +
|\varphi|^2)^2}$ and \cite{Nitta:2014pwa}
\begin{align}
\Lambda = (\partial_m \Phi \partial^m \Phi \partial_n
 \Phi^{\dagger} \partial^n \Phi^{\dagger})^{-1} \frac{1}{(1 + \Phi
 \Phi^{\dagger})^4} 
\left[
(\partial_m \Phi^{\dagger} \partial^m \Phi)^2 - \partial_m \Phi
 \partial^m \Phi \partial_n \Phi^{\dagger} \partial^n \Phi^{\dagger}
\right].  \label{eq:FSmodel}
\end{align}
This does not contain 
an additional term other than Fadeev-Skyrme term,
in contrast to Refs.~\cite{BeNeSc,Fr} that contain an additional term.
The other examples include a supersymmetric completion of the Galileon inflation
model \cite{KhLeOv}, the ghost condensation \cite{KoLeOv2} 
and the effective action of the supersymmetric Wess-Zumino model and QCD 
\cite{Kuzenko:2014ypa, Gates:1995fx}. 

\paragraph{Non-canonical branch}
Although it is not easy to find explicit solutions $F^i \not= 0$ for
the $n > 1$ case, one finds the solution for a single chiral
superfield model \cite{Nitta:2014pwa}: 
\begin{align}
F = e^{i\eta} \sqrt{- \frac{K_{\varphi \bar{\varphi}}}{2\Lambda} +
  \partial_m \varphi \partial^m \bar{\varphi}},
\label{eq:non-canonical_sol}
\end{align}
where $\eta$ is a phase factor and $K_{\varphi \bar{\varphi}} =
\frac{\partial^2 K}{\partial \varphi \partial \bar{\varphi}}$.
Then the bosonic part of the on-shell Lagrangian in the non-canonical branch is 
\begin{align}
\mathcal{L}_b =& \ \Lambda |\partial_m \varphi \partial^m \varphi|^2 -
 \Lambda (\partial_m \varphi \partial^m \bar{\varphi})^2 -
 \frac{K^2_{\varphi \bar{\varphi}}}{4 \Lambda}.
\label{eq:neutral_non-cano}
\end{align}
In this case, the ordinary canonical (second space-time derivative)
kinetic term cancels out and 
the on-shell Lagrangian contains higher derivative terms only.
An example is the BPS baby Skyrme model \cite{AdQuSaGuWe},
where $\Lambda$ is given by 
\begin{align}
\Lambda = \frac{1}{(1 + \Phi \Phi^{\dagger})^4}.
\label{eq:Lambda_bS}
\end{align}
The K\"ahler metric is the Fubini-Study metric of $\mathbb{C}P^1$.

A few comments are in order for the non-canonical branch.
First, since $F \bar{F} \ge 0$, the fields satisfy the constraint 
\begin{align}
\partial_m \varphi \partial^m \bar{\varphi} - \frac{K_{\varphi
 \bar{\varphi}}}{2 \Lambda} \ge 0.
\label{eq:noncanonical_constraint}
\end{align}
Second, the last term in Eq.~\eqref{eq:neutral_non-cano} is considered as
the scalar potential when $\Lambda$ does not contain space-time derivative term.
One can introduce an arbitrary scalar potential without the superpotential
$W$ or the D-term potential in the non-canonical branch. 
This is an alternative way to introduce the scalar potential in
supersymmetric models \cite{KoLeOv}.

\subsection{Gauged higher-derivative chiral models}
In this subsection we study couplings of the gauge field to the
higher derivative chiral models.
We consider the higher derivative model of the type
\eqref{eq:ungauged_HD_Lagrangian} where some global symmetries are assumed.
Let us consider the chiral superfields $\Phi^{ia} \ (a =1, \ldots,
\mathrm{dim} G)$ belonging to the fundamental representation of global
symmetry group $G$ with an additional flavor index $i$.\footnote{
It is straightforward to generalize the result in this subsection to
other representations. Therefore we consider the fundamental
representation of $G$ for the chiral superfield $\Phi^a$ throughout this
paper.}
Then the fourth derivative term which preserves the global
symmetry $G$ is 
\begin{align}
\frac{1}{16} \int \! d^4 \theta \ \Lambda_{ik\bar{j} \bar{l}, ab}
 {}^{cd} D^{\alpha} \Phi^{ia} D_{\alpha} \Phi^{k b}
 \bar{D}_{\dot{\alpha}} \Phi^{\dagger \bar{j}}_c \bar{D}^{\dot{\alpha}}
 \Phi^{\dagger \bar{l}}_d,
\label{eq:ungauged_global_4th_derivative}
\end{align}
where the K\"ahler tensor $\Lambda_{ik\bar{j}\bar{l},ab} {}^{cd}$
has indices of the (anti)fundamental representation of $G$.

The gauge field is introduced by the $\mathcal{N} = 1$ vector superfield
$V$ with gauge group $G$. 
The generators $T^{\hat{a}} \ (\hat{a} = 0, 1, \ldots, \text{dim}
\mathcal{G} - 1)$ of the gauge algebra $\mathcal{G}$ are normalized as 
$\mathrm{Tr} [T^{\hat{a}} T^{\hat{b}}] = k \delta^{\hat{a} \hat{b}} \ (k>0)$.
The component expansion of $V = V^{\hat{a}} T^{\hat{a}}$ in the Wess-Zumino gauge is 
\begin{align}
V = - (\theta \sigma^m \bar{\theta}) A_m (x) + i \theta^2 \bar{\theta}
 \bar{\lambda} (x) - i \bar{\theta}^2 \theta \lambda (x) + \frac{1}{2} \theta^2
 \bar{\theta}^2 D (x).
\label{eq:Vector_component}
\end{align}
Here, $A_m$ is the gauge field, $\lambda_{\alpha},
\bar{\lambda}_{\dot{\alpha}}$ are the gauginos, and $D$ is the auxiliary real scalar field.
All the fields belong to the adjoint representation of $G$.
The coupling of the gauge field to the higher derivative terms is
introduced by gauge covariantizing the supercovariant derivatives in
Eq.~\eqref{eq:ungauged_global_4th_derivative}.
The gauge covariantized supercovariant derivative is defined
by 
\begin{align}
\mathcal{D}_{\alpha} \Phi^{ia} = D_{\alpha} \Phi^{ia} + (\Gamma_{\alpha})^a {}_b \Phi^{ib}.
\end{align}
Here $\Gamma_{\alpha}$ is the gauge connection defined by 
\begin{align}
\Gamma_{\alpha} = e^{-2gV} D_{\alpha} e^{2gV},
\end{align}
where $g$ is the gauge coupling constant.
The gauge transformations of the superfields are 
\begin{align}
\Phi^i \to e^{-i \Theta} \Phi^i, \qquad e^{2gV} \to e^{- i
 \Theta^{\dagger}} e^{2gV} e^{i \Theta},
\end{align}
where $\Theta = \Theta^{\hat{a}} (x,\theta,\bar\theta)T^{\hat{a}}$ is 
a gauge parameter chiral superfield.
Then the quantities $\mathcal{D}_{\alpha} \Phi^i,
\bar{\mathcal{D}}_{\dot{\alpha}} \Phi^{\dagger \bar{i}}$ are transformed covariantly
under the gauge transformation:
\begin{align}
\mathcal{D}_{\alpha} \Phi^i \to e^{- i \Theta} \mathcal{D}_{\alpha} \Phi^i, 
\qquad 
\bar{\mathcal{D}}_{\dot{\alpha}} \Phi^{\dagger \bar{i}} \to 
\bar{\mathcal{D}}_{\dot{\alpha}} \Phi^{\dagger \bar{i}} e^{i \Theta^{\dagger}}.
\end{align}
We note that the K\"ahler tensor $\Lambda_{ik\bar{j}\bar{l},ab} {}^{cd}$
becomes a function of $\Phi,\Phi^{\dagger}$ and $V$ in general.

Now we look for the concrete realizations of the 
gauge invariant generalization of the higher
derivative term \eqref{eq:ungauged_global_4th_derivative}.
We find a manifestly gauge invariant generalization of
\eqref{eq:ungauged_global_4th_derivative} is given by 
\begin{align}
- \frac{1}{16} 
\int \! d^4 \theta \ \Lambda_{ik \bar{j} \bar{l}}
 (\Phi,\Phi^{\dagger}, V) (\bar{\mathcal{D}}_{\dot{\alpha}}
 \Phi^{\dagger \bar{j}} e^{2gV} \mathcal{D}^{\alpha} \Phi^i)
 (\bar{\mathcal{D}}^{\dot{\alpha}} \Phi^{\dagger \bar{l}} e^{2gV}
 \mathcal{D}_{\alpha} \Phi^{k}),
\label{eq:4th_deri_gauge2}
\end{align}
where the K\"ahler tensor is
\begin{align}
\Lambda_{ik\bar{j}\bar{l} ab} {}^{cd} = \Lambda_{ik\bar{j} \bar{l}} (\Phi, \Phi^{\dagger}, V) (e^{2gV})^c
 {}_a (e^{2gV})^d {}_b
\end{align}
and $\Lambda_{ik\bar{j}\bar{l}}$ is a gauge invariant $(2,2)$ K\"ahler
tensor which is a function of $\Phi,\Phi^{\dagger}, V$.

The component expansion of the fourth derivative term 
\eqref{eq:4th_deri_gauge2} is 
\begin{align}
& - \frac{1}{16} (\bar{\mathcal{D}}_{\dot{\alpha}} \Phi^{\dagger \bar{j}} e^{2gV} \mathcal{D}^{\alpha} \Phi^i)
(\bar{\mathcal{D}}^{\dot{\alpha}} \Phi^{\dagger \bar{l}} e^{2gV}
 \mathcal{D}_{\alpha} \Phi^k)
\notag \\
=& \ \theta^2 \bar{\theta}^2 
\left[
\frac{}{}
(D^m \bar{\varphi}_a^{\bar{j}} D^n \varphi^{ia}) (D_m \bar{\varphi}_b^{\bar{l}} D_n \varphi^{kb}) 
- \frac{1}{2} (D_m \varphi^{ia} F^{kb} + F^{ia} D_m \varphi^{kb}) (D^m
 \bar{\varphi}_a^{\bar{j}} \bar{F}_b^{\bar{l}} + \bar{F}_a^{\bar{j}} D^m \bar{\varphi}_b^{\bar{l}}
 )
\right.
\notag \\
& \ \qquad \qquad 
\left. 
\frac{}{}
+ F^{ia} \bar{F}^{\bar{j}}_a F^{kb} \bar{F}^{\bar{l}}_b
\right] + I'_f,
\label{eq:4th_deri_gauge2_component}
\end{align}
where $I_f'$ is terms that contain fermions.
Again, there are no auxiliary fields with space-time derivatives in the
purely bosonic terms.
Since the bosonic terms in $\bar{\mathcal{D}}_{\dot{\alpha}}
\Phi^{\dagger} \mathcal{D}^{\alpha} \Phi
\bar{\mathcal{D}}^{\dot{\alpha}} \Phi^{\dagger} \mathcal{D}_{\alpha}
\Phi$ already saturate the Grassmann coordinate, the factor $e^{2gV}$ does not
contribute to the purely bosonic sector of the
Lagrangian. However, the factor $e^{2gV}$ is necessary for the gauge
invariance of the higher derivative terms and this indeed contributes to the fermionic
part $I'_f$ in Eq.~\eqref{eq:4th_deri_gauge2_component}.
We also note that the lowest components in $\Lambda_{ik\bar{j}\bar{l}}$ come from the
chiral superfields only. This is because the lowest component in the
vector superfield $V$ contains the Grassmann coordinate $\theta$ in the Wess-Zumino gauge
\eqref{eq:Vector_component}.
In Ref.~\cite{AdQuSaGuWe}, a three-dimensional analogue of the gauge invariant higher derivative model for a $U(1)$ gauge group was discussed.

Introducing the ordinary kinetic terms for $\Phi^{ia}$ and the gauge field, 
the total Lagrangian we consider is
\begin{align}
\mathcal{L} =& \ \int \! d^4 \theta \ K(\Phi^{\dagger}, \Phi, V) -
 \frac{1}{16} \int \! d^4 \theta \ \Lambda_{ik\bar{j}\bar{l}} (\Phi, \Phi^{\dagger}, V) 
(\bar{\mathcal{D}}_{\dot{\alpha}} \Phi^{\dagger \bar{j}} e^{2gV}
 \mathcal{D}^{\alpha} \Phi^i) (\bar{\mathcal{D}}^{\dot{\alpha}}
 \Phi^{\dagger \bar{l}} e^{2gV} \mathcal{D}_{\alpha} \Phi^k) 
\notag \\
& + \frac{1}{16 k g^2} \mathrm{Tr} 
\left[
\int \! d^2 \theta \ W^{\alpha} W_{\alpha} + ({\rm h.c.})
\right]
- 2 \kappa g \int \! d^4 \theta \ \mathrm{Tr} V.
\label{eq:gauged_model}
\end{align}
Here we have introduced the Fayet-Iliopoulos parameter $\kappa$ 
for the purpose of later discussions.
The field strength of the vector superfield $V$ is defined by
\begin{align}
W_{\alpha} = - \frac{1}{4} \bar{D}^2 (e^{-2gV} D_{\alpha} e^{2gV}).
\end{align}
Throughout this paper, we consider the gauge invariant K\"ahler potential
of the form $K (\Phi^{\dagger}, \Phi, V) = \frac{1}{2} (K(\Phi^{\dagger} e^{2gV},
\Phi) + K(\Phi^{\dagger}, e^{2gV} \Phi))$ and general gauge group $G$ if
not mentioned.
Then, the bosonic component of the Lagrangian \eqref{eq:gauged_model} is 
\begin{align}
\mathcal{L}_{b} =& \ 
- \frac{\partial^2 K}{\partial \bar{\varphi}_a^{\bar{j}} \partial \varphi^{ib}} 
D_m \bar{\varphi}_a^{\bar{j}} D^m \varphi^{ib}
- \frac{\partial^2 K}{\partial \bar{\varphi}_a^{\bar{j}} \partial \varphi^{ib}}
\bar{F}_a^{\bar{j}} F^{ib} + 
\frac{g}{2} D^{\hat{a}}
\left(
\bar{\varphi}_c^{\bar{j}} (T^{\hat{a}})^c {}_d \frac{\partial K}{\partial \bar{\varphi}_d^{\bar{j}}}
+
\frac{\partial K}{\partial \varphi^{ic}} (T^{\hat{a}})^c
 {}_d \varphi^{id} - 2 \kappa \delta^{\hat{a}} {}_0
\right)
\notag \\
& \ + \frac{1}{k} \mathrm{Tr} 
\left[
- \frac{1}{4} F_{mn} F^{mn} + \frac{1}{2} D^2
\right] 
\notag \\
& \ + \Lambda_{ik\bar{j}\bar{l}} (\varphi, \bar{\varphi}) 
\left[
\frac{}{}
(D^m \bar{\varphi}_a^{\bar{j}} D^n \varphi^{ia}) (D_m \bar{\varphi}_b^{\bar{l}} D_n \varphi^{kb}) 
\right.
\notag \\
& \ \qquad \qquad \qquad 
\left. 
\frac{}{}
- \frac{1}{2} (D_m \varphi^{ia} F^{kb} + F^{ia} D_m \varphi^{kb}) (D^m
 \bar{\varphi}_a^{\bar{j}} \bar{F}_b^{\bar{l}} + \bar{F}_a^{\bar{j}} D^m \bar{\varphi}_b^{\bar{l}}
 )
+ F^{ia} \bar{F}^{\bar{j}}_a F^{kb} \bar{F}^{\bar{l}}_b
\right],
\end{align}
where we have assigned the $U(1)$ generator to $T^0$.
The gauge field strength is 
\begin{align}
F_{mn} = \partial_m A_n - \partial_n A_m + i g [A_m, A_n].
\end{align}
The equation of motion for the auxiliary field $D$ is\footnote{
We never introduce higher derivative terms of the vector
superfield $V$. Therefore the equation of motion for $D$ is always
linear and can be solved trivially.} 
\begin{align}
D^{\hat{a}} + \frac{g}{2}
\left(
 \bar{\varphi}_c^{\bar{j}} (T^{\hat{a}})^c {}_d \frac{\partial K}{\partial
 \bar{\varphi}_d^{\bar{j}}} 
+ \frac{\partial K}{\partial \varphi^{ic}} (T^{\hat{a}})^c {}_d \varphi^{id} 
\right)
- g \kappa \delta^{\hat{a}} {}_0 = 0.
\end{align}
The equation of motion for $\bar{F}_a^{\bar{j}}$ is 
\begin{align}
\frac{\partial^2 K}{\partial \bar{\varphi}_a^{\bar{j}} \partial
 \varphi^{ib}} F^{ib} 
-  \Lambda_{ik\bar{j}\bar{l}} (\varphi, \bar{\varphi}) 
\left[
D_m \varphi^{ib} D^m \bar{\varphi}_b^{\bar{j}} F^{ka} + D_m \varphi^{ia}
 D^m \bar{\varphi}^{\bar{l}}_b F^{kb} - 2 F^{ia} F^{kb} \bar{F}^{\bar{l}}_b 
\right]
 = 0.
\label{eq:auxiliary_eom_gauged}
\end{align}
As in the case of the non-gauged chiral superfield models, 
there are two on-shell branches associated with solutions to the equation
\eqref{eq:auxiliary_eom_gauged}. 

\paragraph{Canonical branch}
We first consider the canonical branch.
One finds that $F^{ia} = 0$ is always a solution.
Then, the on-shell Lagrangian in the canonical branch is 
\begin{align}
\mathcal{L}_{b} =& \ 
- \frac{\partial^2 K}{\partial \bar{\varphi}_a^{\bar{j}} \partial \varphi^{ib}} 
D_m \bar{\varphi}^{\bar{j}}_a D^m \varphi^{ib}
+ \Lambda_{ik\bar{j}\bar{l}} (\varphi, \bar{\varphi}) (D^m \bar{\varphi}_a^{\bar{j}} D^n \varphi^{ia})
 (D_m \bar{\varphi}_b^{\bar{l}} D_n \varphi^{kb})
\notag \\
& \ - \frac{g^2}{2} 
\left(
\frac{1}{2} \bar{\varphi}_c^{\bar{j}} (T^{\hat{a}})^c {}_d \frac{\partial K}{\partial
 \bar{\varphi}_d^{\bar{j}}}
+ \frac{1}{2} \frac{\partial K}{\partial \varphi^{ic}} (T^{\hat{a}})^c {}_d \varphi^{id}
 - \kappa \delta^{\hat{a}} {}_0
\right)^2 - \frac{1}{4k} \mathrm{Tr} F_{mn} F^{mn}. 
\label{eq:canonical_Lag}
\end{align}
The vacuum of the model is determined by the D-term condition
\begin{align}
\bar{\varphi}_c^{\bar{i}} (T^{\hat{a}})^c {}_d \varphi^{id} - \kappa \delta^a {}_0 = 0.
\end{align}
We stress that $\Lambda_{ik\bar{j}\bar{l}}$ 
does not contain
the space-time derivatives on $\Phi$ ($\Phi^\dagger$), 
unlike the non-gauged cases 
for which the space-time derivative can act on 
$\Phi$ ($\Phi^\dagger$) in $\Lambda_{ik\bar{j}\bar{l}}$.
This is because the gauge covariant derivative of a chiral superfield $D_m
\Phi^{ia}$ does not provide supersymmetric couplings of the gauge
field. From now on, we therefore consider the tensor $\Lambda_{ik\bar{j}\bar{l}}$
which never contains the space-time derivatives of the superfields.

\paragraph{Non-canonical branch}
It is not so easy to find a $F^{ia} \not= 0$ solution even
for the single chiral superfield model.
However, we find that a $F^a \not= 0$ solution can be explicitly written
down for single chiral superfield models with a $U(1)$ gauge group as 
\begin{align}
F^0 = 
e^{i \eta} \sqrt{
 - \frac{K_{\varphi \bar{\varphi}}}{2 \Lambda} + D_m \varphi D^m
 \bar{\varphi}
},
\label{eq:non-canonical_sol_gauged}
\end{align}
where $\eta$ is a phase factor.
The solution in Eq.~\eqref{eq:non-canonical_sol_gauged} is just the gauge
covariantized counterpart of that in Eq.~\eqref{eq:non-canonical_sol}.
The fields satisfy the gauge covariantized constraint \eqref{eq:noncanonical_constraint}.
\begin{align}
|F^0|^2 = 
 - \frac{K_{\varphi \bar{\varphi}}}{2 \Lambda} + D_m \varphi D^m
 \bar{\varphi} \ge 0.
\label{eq:noncanonical_constraint_gauged}
\end{align}
Then the bosonic part of the on-shell Lagrangian in the non-canonical
branch is 
\begin{align}
\mathcal{L}_{b} =& - \frac{1}{4} F_{mn} F^{mn} 
- \frac{g^2}{2} 
\left(
\frac{1}{2} \bar{\varphi} \frac{\partial K}{\partial \bar{\varphi}}
+ \frac{1}{2} \frac{\partial K}{\partial \varphi} \varphi
 - \kappa
\right)^2
\notag \\
& \ + \Lambda (|D_m \varphi D^m \varphi|^2 - (D_m \varphi D^m \bar{\varphi})^2)
- \frac{(K_{\varphi \bar{\varphi}})^2}{4 \Lambda},
\label{eq:non-canonical_Lag}
\end{align}
where $F_{mn} = \partial_m A_n - \partial_n A_m$ is the field strength
of the $U(1)$ gauge field.
An example of the Lagrangian \eqref{eq:non-canonical_Lag} is a
supersymmetric generalization of the gauged BPS baby Skyrme model
\cite{Adam:2012pm} whose potential term is determined by the K\"ahler
potential $K$ through the D-term and the term $K^2_{\varphi
\bar{\varphi}}/\Lambda$.
In this case, the explicit function $\Lambda$ is given 
in Eq.~\eqref{eq:Lambda_bS}.

\section{BPS states in Minkowski space}\label{sec:BPS-Minkowski}
In this section, we investigate BPS configurations of the model
\eqref{eq:gauged_model} in four-dimensional Minkowski space.
BPS configurations in supersymmetric theories preserve parts of
supersymmetry. BPS equations are obtained from the condition that the
on-shell supersymmetry transformation of the fermions in the model
vanishes, $\delta^{\text{on}}_{\xi} \psi_{\alpha} =
\delta^{\text{on}}_{\xi} \lambda_{\alpha} = 0$.
Here $\delta^{\text{on}}_{\xi}$ ($\delta^{\text{off}}_{\xi}$) is the on-shell
(off-shell) supersymmetry transformation by the parameters
$\xi_{\alpha}$, $\bar{\xi}^{\dot{\alpha}}$.
The off-shell supersymmetry variation of the fermions $\psi$, $\lambda$ is
\begin{align}
\delta^{\text{off}}_{\xi} \psi_{\alpha}^{ia} =& \ \sqrt{2} i
 (\sigma^m)_{\alpha \dot{\alpha}} \bar{\xi}^{\dot{\alpha}} D_m
 \varphi^{ia} + \sqrt{2} \xi_{\alpha} F^{ia}, \\
\delta^{\text{off}}_{\xi} \lambda_{\alpha} =& \ 
i \xi_{\alpha} D + (\sigma^{mn})_{\alpha} {}^{\beta} \xi_{\beta}
 F_{mn}.
\end{align}
The on-shell supersymmetry transformations are obtained by substituting
the solutions of the auxiliary fields equations into $F$ and $D$.
Therefore they have distinct structures in the canonical and
non-canonical branches.

In Ref.~\cite{Nitta:2014pwa}, we studied BPS equations in the non-gauged higher
derivative models given in Eq.~\eqref{eq:off-shell_Lagrangian} where no gauge fields are present.
We derived the 1/2 BPS domain wall and lump equations in the canonical branch.
These equations are the same for the ordinary (without higher derivative
term) theory. We calculated the BPS bound of the on-shell action  
associated with these configurations. Then we found that the BPS bound is
given by the ordinary tension of the domain wall and the lump
(topological) charge, respectively. 
Namely, higher derivative effects are totally
canceled in the 1/2 BPS domain wall and lump.
In the non-canonical branch, we found 1/4 BPS configurations for the
domain wall junctions and lump type solitons.
The equation for the domain wall junction receives higher derivative
contributions while the associated BPS bound of the Lagrangian is expressed by the ordinary domain
wall tension and the junction charge.
For the lump type soliton, it is considered as a compacton which is a
soliton with a compact support. Indeed, when the K\"ahler potential $K$ and
$\Lambda$ are chosen appropriately, the 1/4 BPS equation in Ref.~\cite{Nitta:2014pwa} have compacton type solutions \cite{AdQuSaGuWe}.

In the following subsections, we proceed with the analysis of the BPS
configurations for the gauged higher derivative chiral models given in Eq.~\eqref{eq:gauged_model}.
For the ordinary $\mathcal{N} = 1$ supersymmetric gauge theory with fundamental
matters in Minkowski space, there are BPS vortices which are codimension
two solitons. 
We study codimension-two vortex configurations
in the canonical and non-canonical branches of the model \eqref{eq:gauged_model}.

\subsection{Canonical branch}
We start from the flat K\"ahler potential $K = \Phi^{\dagger \bar{i}} e^{2gV}
\Phi^i$ and look for the vortex configurations.
The static ansatz for the vortex is given by 
\begin{align}
\varphi^{ia} = \varphi^{ia} (x^1,x^2), \qquad F_{12} \not= 0,
\end{align}
where the other components of $F_{mn}$ all vanish.
In the canonical branch, we have the solution $F^{ia} = 0$.
Then, the on-shell supersymmetry variations of the fermions are 
\begin{align}
\delta \psi^i =& \  \sqrt{2}i 
\left(
\begin{array}{cc}
(D_1 - i D_2) \varphi^i \bar{\xi}^{\dot{2}} \\
(D_1 + i D_2) \varphi^i \bar{\xi}^{\dot{1}}
\end{array}
\right) = 0, \\ 
\delta \lambda =& \ 
 - i 
\left(
\begin{array}{c}
\xi_1 F_{12} - \xi_1 D\\
- \xi_2 F_{12}  - \xi_2 D
\end{array}
\right) = 0,
\end{align}
where $D^{\hat{a}} = - g
\left(
\bar{\varphi}_c^{\bar{i}} (T^{\hat{a}})^c {}_d \varphi^{id}
- \kappa \delta^{\hat{a}} {}_0
\right)$.
The vortex configuration is obtained by imposing the following 
projection condition on the supersymmetry parameter:
\begin{align}
\frac{1}{2} (\sigma^1 + i \sigma^2) \bar{\xi} = 0.
\label{eq:half_projection}
\end{align}
This is equivalent to the condition $\bar{\xi}^{\dot{2}} = \xi_1 = 0$ so
that the projection \eqref{eq:half_projection} leaves a half of 
$\mathcal{N} = 1$ supersymmetry.
Therefore, we obtain the following BPS equations: 
\begin{align}
\bar{D}_z \varphi^{ia} = 0, \qquad F_{12}^{\hat{a}} - g
\left(
\bar{\varphi}_c^{\bar{i}} (T^{\hat{a}})^c {}_d \varphi^{id} - \kappa \delta^{\hat{a}} {}_0
\right) = 0.
\label{eq:canonical_half_BPS_vortex}
\end{align}
Here we have defined $z \equiv \frac{1}{2} (x^1 + i x^2)$ and 
$D_z \equiv D_1 - i D_2$, $\bar{D}_z \equiv D_1 + i D_2$.
This is just the ordinary 1/2 BPS Abelian (ANO) or non-Abelian vortex equation 
\cite{Hanany:2003hp}.
Now we calculate the Lagrangian bound\footnote{When the Lagrangian \eqref{eq:canonical_Lag}
contains higher order time derivatives of fields, the positive energy Hamiltonian
is not defined in general \cite{Ostrogradski}. Therefore, we calculate the Lagrangian bound,
rather than the energy bound, for the BPS configurations.} 
associated with the BPS equations \eqref{eq:canonical_half_BPS_vortex}. Using the first
condition in Eq.~\eqref{eq:canonical_half_BPS_vortex}, we find the higher
derivative terms vanish:
\begin{align}
& \Lambda_{ik\bar{j}\bar{l}} 
(D^m \bar{\varphi}_a^{\bar{j}} D^n \varphi^{ia}) (D_m \bar{\varphi}_b^{\bar{l}} D_n
 \varphi^{kb}) 
\notag \\
=& \ \frac{1}{4} \Lambda_{ik\bar{j}\bar{l}}
\left(
D_z \varphi^{ia} \bar{D}_z \varphi^{kb} + \bar{D}_z \varphi^{ia} D_z \varphi^{kb}
\right)
\left(
D_z \bar{\varphi}_a^{\bar{j}} \bar{D}_z \bar{\varphi}_b^{\bar{l}} + \bar{D}_z
 \bar{\varphi}_a^{\bar{j}} D_z \bar{\varphi}_b^{\bar{l}}
\right) 
\notag \\
=& \  0.
\end{align}
Then, by using the first and the second equations in 
\eqref{eq:canonical_half_BPS_vortex}, 
we obtain the Lagrangian bound 
\begin{align}
\mathcal{L} =& \  \kappa g F^0_{12}.
\label{eq:canonical_vortex_bound}
\end{align}
Here $F^0_{12}$ is the $U(1)$ flux density in the $(x^1,x^2)$-plane. 
Integrating it in the $(x^1,x^2)$-plane, we obtain the ordinary vortex
topological charge. Therefore, in the canonical branch, 
all the higher derivative corrections to the
1/2 BPS vortex are canceled in both the equations
\eqref{eq:canonical_half_BPS_vortex} and the Lagrangian bound
\eqref{eq:canonical_vortex_bound}.
This is a conceivable result since the BPS nature is determined by the
supersymmetry algebra. The model \eqref{eq:canonical_Lag} includes
higher derivative terms but supersymmetry is manifestly realized.
Then we expect that the BPS structure is protected against higher
derivative corrections.
A typical example is the world-volume theory of D-branes where BPS states in
super Yang-Mills theory linearize the non-Abelian DBI action canceling the higher derivative corrections 
\cite{Brecher:1998tv}.
While the higher derivative corrections exist in 
the non-Abelian vortex effective theory,
the higher derivative effects are canceled in the BPS equation and energy of ${\mathbb C}P^{N-1}$ lumps 
inside a non-Abelian vortex 
\cite{Eto:2012qda}.
We also comment that this is the same conclusion discussed in
the domain wall and lump in the non-gauged chiral models \cite{Nitta:2014pwa}.

We next consider the general gauge invariant K\"ahler potential of the
form $K(\Phi^{\dagger}, \Phi, V) = \frac{1}{2} (K(\Phi^{\dagger}
e^{2gV}, \Phi) + K(\Phi^{\dagger}, e^{2gV} \Phi))$.
The BPS equations for the 1/2 BPS projection condition \eqref{eq:half_projection}
are  
\begin{align}
\bar{D}_z \varphi^{ia} = 0, \qquad F_{12}^{\hat{a}} - \frac{g}{2} 
\left(
\bar{\varphi}_c^{\bar{j}} (T^{\hat{a}})^c {}_d \frac{\partial K}{\partial \bar{\varphi}_d^{\bar{j}}}
 + 
\frac{\partial K}{\partial \varphi^{ic}} (T^{\hat{a}})^c {}_d \varphi^{id}
- \kappa \delta^{\hat{a}} {}_0
\right) = 0.
\label{eq:canonical_half_BPS_general_K}
\end{align}
By using the first condition in
\eqref{eq:canonical_half_BPS_general_K}, we find that the higher derivative terms
vanish. Then, the Lagrangian bound associated with the BPS condition
\eqref{eq:canonical_half_BPS_general_K} is 
\begin{align}
\mathcal{L} =& \ - \frac{1}{2} \frac{\partial^2 K}{\partial
 \bar{\varphi}_a^{\bar{j}} \partial \varphi^{ib}}
\bar{D}_z \bar{\varphi}_a^{\bar{j}} D_z \varphi^{ib} -  
\frac{g^2}{2} 
\left(
\frac{1}{2} \bar{\varphi}_c^{\bar{j}} (T^{\hat{a}})^c {}_d \frac{\partial K}{\partial
 \bar{\varphi}_d^{\bar{j}}}
+ \frac{1}{2} \frac{\partial K}{\partial \varphi^{ic}} (T^{\hat{a}})^c {}_d
 \varphi^{id} - \kappa \delta^{\hat{a}} {}_0
\right)^2 
- \frac{1}{2} (F^{\hat{a}}_{12})^2 
\notag \\
=& \ 
- \varepsilon^{st} \partial_s \mathcal{N}_t + \kappa g
 F^0_{12},
\label{eq:canonical_bound2}
\end{align}
where we have defined the following quantity 
\begin{align}
\mathcal{N}_s = 
\frac{i}{2} 
\left(
\frac{\partial K}{\partial \bar{\varphi}_a^{\bar{j}}} D_s \bar{\varphi}_a^{\bar{j}} -
 \frac{\partial K}{\partial \varphi^{ia}} D_s \varphi^{ia}
\right), \qquad (s,t = 1,2).
\end{align}
The first term in Eq.~\eqref{eq:canonical_bound2} is the gauge covariant
generalization of the lump charge density.
Then the Lagrangian bound is given by the sum of the lump and the
vortex charge densities.
The BPS configurations whose energy bound is given by
Eq.~\eqref{eq:canonical_bound2} have been studied in the gauged non-linear
sigma models where higher derivative corrections are absent \cite{Schroers:1995he,Nitta:2011um}.
In there, the configurations admit fractional lump charges.
Once again, we find that all the higher derivative effects are canceled on the 1/2
BPS states \eqref{eq:canonical_half_BPS_general_K}.

\subsection{Non-canonical branch}
We next consider BPS equations in the non-canonical branch.
The Lagrangian is given by \eqref{eq:non-canonical_Lag} where the gauge
group is $U(1)$ and $K = \Phi^{\dagger} e^{2gV} \Phi$.
The non-zero solution of the auxiliary field $F^0$ is given in Eq.~\eqref{eq:non-canonical_sol_gauged}.
The supersymmetry variation of the fermions is 
\begin{align}
\delta \psi =& \  \sqrt{2} 
\left(
\begin{array}{cc}
i (D_1 - i D_2) \varphi \bar{\xi}^{\dot{2}} + \xi_1 F^0
\\
i (D_1 + i D_2) \varphi \bar{\xi}^{\dot{1}} + \xi_2 F^0
\end{array}
\right) = 0, \\ 
\delta \lambda =& \ 
 - i 
\left(
\begin{array}{c}
\xi_1 F_{12} - \xi_1 D \\
- \xi_2 F_{12} - \xi_2 D
\end{array}
\right)
= 0.
\end{align}
Since the auxiliary field $F^0$ is non-zero in the non-canonical branch, 
the 1/2 BPS projection \eqref{eq:half_projection} gives the equations 
\eqref{eq:canonical_half_BPS_vortex} and the following additional condition:
\begin{align}
F^0 = e^{i \eta} 
\sqrt{
 - \frac{1}{2 \Lambda} + D_m \varphi D^m
 \bar{\varphi} } = 0.
\label{eq:noncanonical_F_condition}
\end{align}
Solutions that satisfy the ordinary vortex equations
\eqref{eq:canonical_half_BPS_vortex} do not satisfy the condition in Eq.~\eqref{eq:noncanonical_F_condition} for general $\Lambda$.\footnote{However, when $\Lambda$ is chosen appropriately, it is
possible that the ordinary vortex solution satisfies the condition
\eqref{eq:noncanonical_F_condition}.
}
We therefore look for another BPS condition.
A natural candidate is the gauge covariantized generalization of the BPS lumps 
in the non-canonical branch.
Following the BPS lumps studied in Ref.~\cite{Nitta:2014pwa}, we consider the 1/4 BPS projection conditions,
\begin{align}
\frac{1}{2} (\sigma^1 + i \sigma^2)_{\alpha \dot{\alpha}}
 \bar{\xi}^{\dot{\alpha}} = 0, \qquad 
\frac{1}{2} (\sigma^1 - i \sigma^2)_{\alpha \dot{\alpha}}
 \bar{\xi}^{\dot{\alpha}} = i \xi_{\alpha}.
\end{align}
Then, from the variation of the fermions, we find 
a set of 1/4 BPS equations:
\begin{align}
\bar{D}_z \varphi =- i e^{ i \eta} 
\sqrt{
- \frac{1}{2 \Lambda} + \frac{1}{2} 
(D_z \bar{\varphi} \bar{D}_z \varphi + \bar{D}_z \bar{\varphi} D_z \varphi)
}, \qquad 
F^0_{12} - g 
(\bar{\varphi} \varphi - \kappa) = 0.
\label{eq:gauged_compacton_flat}
\end{align}
\eqref{eq:noncanonical_constraint_gauged}.
The first equation is the gauge covariantized generalization of the compacton-type equation while the second equation is that for the ANO vortex. 
We call  solutions to these equations as higher derivative vortices. 
These equations may admit a vortex with a compact 
support for the scalar fields (that we may call a compact vortex).
See Ref.~\cite{Adam:2008rf} for a vortex with a compact support 
which are non-BPS in non-supersymmetric theories.

We then calculate the Lagrangian bound associated with the BPS condition
\eqref{eq:gauged_compacton}. 
Using the first condition in Eq.~\eqref{eq:gauged_compacton}, we obtain the following relation,
\begin{align}
\Lambda 
\left\{
(D_m \varphi D^m \varphi) (D_n \bar{\varphi} D^n \bar{\varphi}) - (D_m
 \varphi D^m \bar{\varphi})^2
\right\} = 
- \frac{1}{4} \Lambda 
\left(
\bar{D}_z \varphi D_z \bar{\varphi} - D_z \varphi \bar{D}_z \bar{\varphi}
\right)^2 = - \frac{1}{4\Lambda}.
\end{align}
By using this relation and the second equation in Eq.~\eqref{eq:gauged_compacton_flat}, 
we calculate the BPS bound of the Lagrangian as 
\begin{align}
\mathcal{L} = \kappa g F^0_{12}.
\end{align}
This is the topological vortex charge density.
Therefore the equations \eqref{eq:gauged_compacton_flat} correspond to
the higher derivative generalization of the ANO vortex rather than the compacton.
We comment that the higher derivative terms cancel out in the Lagrangian
bound even in the non-canonical branch. However, the BPS equation
\eqref{eq:gauged_compacton} receives higher derivative corrections.
The situation is quite similar to the 1/4 BPS domain wall junction and
the compacton in the non-gauged model \cite{Nitta:2014pwa}. 
In there, there are higher derivative corrections to the BPS equations. 
However, the bounds for the BPS states do not receive higher derivative corrections.

Now we consider the general gauge invariant K\"ahler potential.
A set of 1/4 BPS equations is obtained as 
\begin{align}
\bar{D}_z \varphi =- i e^{ i \eta} 
\sqrt{
- \frac{K_{\varphi \bar{\varphi}}}{2 \Lambda} + \frac{1}{2} 
(D_z \bar{\varphi} \bar{D}_z \varphi + \bar{D}_z \bar{\varphi} D_z \varphi)
}, \qquad 
F^0_{12} - \frac{g}{2} 
\left(\bar{\varphi} \frac{\partial K}{\partial \bar{\varphi}} 
+ \frac{\partial K}{\partial \varphi} \varphi
- \kappa\right) = 0.
\label{eq:gauged_compacton}
\end{align}
Using the first condition in Eq.~\eqref{eq:gauged_compacton}, we find that
 the higher derivative terms cancel out in the Lagrangian bound. 
The result is 
\begin{align}
\mathcal{L} =  - \varepsilon^{st} \partial_s \mathcal{N}_t + \kappa g
 F^0_{12}, \quad (s,t = 1,2),
\end{align}
where 
\begin{align}
\mathcal{N}_s = \frac{i}{2} (K_{\bar{\varphi}} D_s \bar{\varphi} -
 K_{\varphi} D_s \varphi).
\end{align}
This is precisely the sum of the lump and the vortex charges.
We therefore expect that the equations \eqref{eq:gauged_compacton}
describe composite states of the higher derivative ANO vortex and 
the BPS baby Skyrmions, 
or simply gauged BPS baby Skyrmions.  
Solutions should carry fractional baby Skyrmion charges 
as for the vortex-lumps in the canonical branch.
BPS states in Minkowski space are summarized in Table.~\ref{tb:HDBPS_Minkowski}

\begin{table}[tb]
\begin{center}
\begin{tabular}{|l||l|l|l|}
\hline
 & SYM + SUSY NLSM & canonical & non-canonical \\
\hline \hline
L type & 1/2 BPS lump & 1/2 BPS lump & 1/4 BPS baby-Skyrmion \\
\hline
V type & 1/2 BPS vortex & 1/2 BPS vortex & 1/4 BPS HD vortex \\
\hline
VL type & 1/2 BPS vortex-lump & 1/2 BPS vortex-lump & 1/4 BPS vortex-baby Skyrmion \\
\hline
\end{tabular}
\caption{BPS states in the gauged higher derivative (HD) chiral model
 and super Yang-Mills with gauged non-linear sigma model (SUSY NLSM). 
Theories are defined in Minkowski space. The BPS states are classified
 into the lump (L) type, the vortex (V) type and the vortex-lump (VL) type.}
\label{tb:HDBPS_Minkowski}
\end{center}
\end{table}

\section{BPS states in Euclidean space}\label{sec:BPS-Euclidean}
In four-dimensional Euclidean space, one can consider codimension-four
objects. Typical examples are the Yang-Mills instantons and the instantons
trapped inside (intersecting) vortices. 
In this section, we study codimension-four BPS configurations of the higher derivative model
\eqref{eq:gauged_model} in Euclidean space. 
The off-shell supersymmetry variations of the fermions in Euclidean space
is 
\begin{align}
\delta_{\xi} \psi_{\alpha}^i =& \ \sqrt{2} i (\sigma^m_{\text{E}})_{\alpha
 \dot{\alpha}} \bar{\xi}^{\dot{\alpha}} D_m \varphi^i + \sqrt{2}
 \xi_{\alpha} F^i, \\
\delta_{\xi} \lambda_{\alpha} =& \ i \xi_{\alpha} D +
 (\sigma^{mn}_{\text{E}})_{\alpha} {}^{\beta} \xi_{\beta} F_{mn}, 
\end{align}
where $m = 1,2,3,4$ and the sigma matrices in the Euclidean space are defined by 
\begin{align}
(\sigma_{\text{E}}^{m})_{\alpha \dot{\alpha}} = (i \vec{\tau}, \mathbf{1}), \quad 
(\bar{\sigma}_{\text{E}}^m)^{\dot{\alpha} \alpha} = (- i \vec{\tau}, \mathbf{1}).
\end{align}
Here, $\vec{\tau}$ are the Pauli matrices.
The explicit supersymmetry variation of the fermions are found in Appendix
\ref{sec:SUSY_variation}. 
We note that in Euclidean space, $\xi^{\alpha}$ and
$\bar{\xi}^{\dot{\alpha}}$ are 
independent from each other and they are not complex conjugate anymore. 
Then it is possible to consider BPS 
projections that drop a chiral half of $\mathcal{N} = 1$ supersymmetry $\xi^{\alpha} =
0$, $\bar{\xi}^{\dot{\alpha}} \not= 0$. 
Indeed, the standard Yang-Mills
instantons exist in our model \eqref{eq:gauged_model}, 
that 
preserve the (anti)chiral half of supersymmetry and 
are 1/2 BPS configurations.
Since BPS states with codimensions less than four in Euclidean space are the same as those 
in Minkowski space, 
discussed in the previous section, 
we focus on codimension-four BPS states 
in the higher
derivative model in the following subsections.

\subsection{Canonical branch}
We start from the Lagrangian \eqref{eq:gauged_model} where the K\"ahler
potential is flat. 
We consider the 1/4 BPS projection condition\footnote{The other combinations, for example, 
$\xi_2 \not= 0, \xi_1 = \bar{\xi}^{\dot{1}} = \bar{\xi}^{\dot{2}} = 0$
and so on give essentially the same form of the BPS equations.}
\begin{align}
\bar{\xi}^{\dot{1}} \not= 0, \quad \bar{\xi}^{\dot{2}} = \xi_1 = \xi_2 =
 0.
\label{eq:qBPS_euclidean}
\end{align}
Then from the supersymmetry variation of the fermions, 
we obtain the following set of 1/4 BPS equations in the canonical branch:
\begin{align}
& \bar{D}_z \varphi^i = \bar{D}_w
 \varphi^i = 0, \quad F^{\hat{a}}_{12} - F^{\hat{a}}_{34} = g (\bar{\varphi}_c^{\bar{i}} (T^{\hat{a}})^c
 {}_d \varphi^{id} - \delta^{\hat{a}} {}_0 \kappa) ,
\notag \\
& F^{\hat{a}}_{13} + F^{\hat{a}}_{24} = F^{\hat{a}}_{14} - F^{\hat{a}}_{23} = 0,
\label{eq:vvi_canonical}
\end{align}
where we have defined complex coordinates and derivatives 
with respect to them by
\begin{align}
& z \equiv \frac{1}{2} (x^1 + i x^2), \qquad w \equiv \frac{1}{2} (x^4 + i x^3), 
\notag \\
& D_z \equiv D_1 - i D_2, \qquad D_w \equiv D_4 - i D_3.
\end{align}
Using the condition $\bar{D}_z \varphi^i = \bar{D}_w \varphi^i = 0$, we find
that the higher derivative terms vanish for the BPS configuration
\eqref{eq:vvi_canonical}, 
\begin{align}
& \Lambda_{ik\bar{j}\bar{l}} 
(D_m \bar{\varphi}_a^{\bar{j}} D^m \bar{\varphi}_b^{\bar{l}})
(D_n \varphi^{ib}  D^n \varphi^{kb}) 
\notag \\
=& \ \frac{1}{4} \Lambda_{ik\bar{j}\bar{l}} 
\left(
D_z \varphi^{ia} \bar{D}_z \varphi^{kb} + \bar{D}_z \varphi^{ia} D_z \varphi^{kb} +
 D_w \varphi^{ia} \bar{D}_w \varphi^{kb} + \bar{D}_w \varphi^{ia} D_w \varphi^{kb}
\right)
\notag \\
& \qquad \times 
\left(D_z \bar{\varphi}_a^{\bar{j}} \bar{D}_z \bar{\varphi}_b^{\bar{l}} + 
\bar{D}_z \bar{\varphi}_a^{\bar{j}} D_z \bar{\varphi}_b^{\bar{l}} + D_w \bar{\varphi}_a^{\bar{j}}
 \bar{D}_w \bar{\varphi}_b^{\bar{l}} + \bar{D}_w \bar{\varphi}_a^{\bar{j}} D_w \bar{\varphi}_b^{\bar{l}}
\right) 
\notag \\
=& \ 0.
\end{align}
Then the BPS bound of the Lagrangian associated with the configuration \eqref{eq:vvi_canonical} is 
\begin{align}
\mathcal{L}_{\text{E}} = - \kappa g (F^0_{12} - F^0_{34}) + \frac{1}{4k}
 \mathrm{Tr} [F_{mn} \tilde{F}^{mn}],
\label{eq:EucLagrangian_bound_canonical}
\end{align}
where $\tilde{F}_{mn} = \frac{1}{2} \varepsilon_{mnpq} F^{pq}$ is the
Hodge dual of the gauge field strength $F_{mn}$.
We note that the sign of the Lagrangian in
Euclidean space is flipped from that in Minkowski space.
The first and the second terms in
\eqref{eq:EucLagrangian_bound_canonical} correspond to the vortex charge densities
in the $(x^1,x^2)$ and $(x^3,x^4)$-planes, respectively. The last term is the instanton charge density. 
Therefore solutions to Eq.~\eqref{eq:vvi_canonical} 
are the Yang-Mills instantons trapped inside 
intersecting vortices. 
A set of these equations were first found in Refs.~\cite{Hanany:2004ea,Eto:2004rz,Eto:2006pg,Fujimori:2008ee} for 
supersymmetric theories with eight supercharges 
without higher derivative terms, 
and  configurations were shown to be 1/4 BPS states 
\cite{Eto:2004rz}.
Solutions can be constructed in terms of the moduli matrix 
\cite{Eto:2006pg} and are mathematically 
characterized in terms of 
amoeba and tropical geometry 
\cite{Fujimori:2008ee}. 

We next consider the general gauge invariant K\"ahler potential. 
In this case, a set of 1/4 BPS equations that we obtain is 
\begin{align}
& \bar{D}_z \varphi^i = \bar{D}_w \varphi^i = 0, 
\quad F^{\hat{a}}_{12} - F^{\hat{a}}_{34} = 
\frac{g}{2} 
\left(
\bar{\varphi}_c^{\bar{j}} (T^{\hat{a}})^c {}_d \frac{\partial K}{\partial \bar{\varphi}_d^{\bar{j}}}
+ \frac{\partial K}{\partial \varphi^{ic}} (T^{\hat{a}})^c {}_d \varphi^{id} - \kappa
\delta^{\hat{a}} {}_0 \right), \notag \\
& F^{\hat{a}}_{13} + F^{\hat{a}}_{24} = F^{\hat{a}}_{14} - F^{\hat{a}}_{23} = 0.
\label{eq:qBPS_canonical_generalK}
\end{align}
Using Eqs.~\eqref{eq:qBPS_canonical_generalK}, 
the BPS bound of the Lagrangian can be  evaluated as 
\begin{align}
\mathcal{L}_{\text{E}} = \varepsilon^{st} \partial_s \mathcal{N}_t -
 \varepsilon^{s' t'} \partial_{s'} \mathcal{N}_{t'} - \kappa g (F^0_{12}
 - F^0_{34}) + \frac{1}{4k} \mathrm{Tr} [F_{mn} \tilde{F}^{mn}],
\end{align}
where $s,t = 1,2$ and $s',t' = 3,4$.
The first and the second terms correspond to the gauge covariantized
extension of the lump charge densities in the $(x^1,x^2)$ and
$(x^3,x^4)$-planes, respectively.
The third and the fourth terms are vortex charge densities in the
$(x^1,x^2)$ and $(x^3,x^4)$-planes, respectively, and the last term is the Yang-Mills
instanton charge density.
Note that when the gauge field vanishes, the configuration corresponds to
the intersecting topological vortex-lumps in the $(x^1,x^2)$- and
$(x^3,x^4)$-planes.

\subsection{Non-canonical branch}
Finally, we consider the non-canonical branch where the gauge group is
$U(1)$. 
The 1/4 BPS configurations in the two-dimensional subspaces are
constructed by the same ways discussed in the Minkowski case.
We now look for codimension-four BPS states.
Since the solution of the auxiliary field is not zero in the non-canonical branch, 
the 1/4 BPS projection \eqref{eq:qBPS_euclidean} gives the BPS equations
\eqref{eq:vvi_canonical} and the additional condition $F^0 = 0$
\eqref{eq:noncanonical_F_condition}. 
As in the case of the Minkowski space,
the solutions to the equations \eqref{eq:vvi_canonical} do not
satisfy the condition \eqref{eq:noncanonical_F_condition} for general
$\Lambda$. Therefore the 1/4 BPS configurations associated with the
projection \eqref{eq:qBPS_euclidean} do not exist in the non-canonical
branch. BPS states in Euclidean space are summarized in Table.~\ref{tb:HDBPS_Euclidean}


\begin{table}[tb]
\begin{center}
\begin{tabular}{|l||l|l|l|}
\hline
 & SYM + SUSY NLSM & canonical & non-canonical \\
\hline \hline
L type & 1/2 BPS $\text{L}_{12}$ & 1/2 BPS $\text{L}_{12}$ & 1/4 BPS
$\text{bS}_{12}$ \\
\hline
V type & 1/2 BPS $\text{V}_{12}$ & 1/2 BPS $\text{V}_{12}$ & 1/4 BPS
			 $\text{HDV}_{12}$ \\
\hline
VL type & 1/2 BPS $\text{VL}_{12}$ & 1/2 BPS $\text{VL}_{12}$ & 1/4 BPS
			 $\text{HDVbS}_{12}$ \\
\hline
V-V-I type & 1/4 BPS $\text{V}_{12}$-$\text{V}_{34}$-I  & 
1/4 BPS $\text{V}_{12}$-$\text{V}_{34}$-I & no \\
\hline
VL-VL-I type & 1/4 BPS $\text{VL}_{12}$-$\text{VL}_{34}$-I & 1/4 BPS $\text{VL}_{12}$-$\text{VL}_{34}$-I & no \\
\hline
L-L type & 1/4 BPS $\text{L}_{12}$-$\text{L}_{34}$ & 1/4 BPS $\text{L}_{12}$-$\text{L}_{34}$ & no \\
\hline
\end{tabular}
\caption{BPS states in the gauged higher derivative (HD) chiral model
 and super Yang-Mills with gauged non-linear sigma model. Theories are
 defined in Euclidean space.
Here L,V,I,VL,HDV, bS, and HDVbS stand for lumps, vortices, instantons,
 vortex-lumps, higher derivative vortices, BPS baby Skyrmions, and
 higher derivative vortex-BPS baby Skyrmions, respectively. 
The subscript stands for subspaces that the soliton is defined.
}
\label{tb:HDBPS_Euclidean}
\end{center}
\end{table}


\section{Summary and discussion} \label{sec:summary}

In this paper, 
we have classified BPS states in ${\cal N}=1$  supersymmetric gauge theories 
coupled with higher derivative chiral models 
in four Minkowski and Euclidean dimensions. 
We have found canonical and non-canonical branches
corresponding to solutions $F=0$ and $F\neq 0$ 
of auxiliary field equations, respectively. 
1/2 BPS states 
in theories without higher derivative terms 
remain 1/2 BPS  in the canonical branch  
and the corresponding BPS states in the non-canonical branch 
are 1/4 BPS states. 
1/4 BPS states  
in theories without higher derivative terms 
remain 1/4 BPS in the canonical branch 
but there are no corresponding BPS states in the non-canonical branch.
We have obtained 
1/2 BPS equations for 
an ANO vortex,  
a non-Abelian vortex, a lump, 
and a vortex-lump in the canonical branch, 
and 1/4 BPS higher derivative generalization of the ANO vortices 
in the non-canonical branch. 
In four Euclidean dimensions, we have obtained 
the 1/4 BPS Yang-Mills instantons trapped inside a non-Abelian vortex, 
and 1/4 BPS intersecting vortices or 
vortex-lump intersections with instanton charges 
in the canonical branch and   
no codimension-four BPS states in the non-canonical branch.

While we have given 
the superfield Lagrangian of gauged multi-component chiral models,
we have been able to obtain on-shell Lagrangian only for the cases of 
a single component because of difficulty solving the equations of motion 
for the auxiliary fields for the multi-component cases.
Obtaining on-shell Lagrangians for gauged or non-gauged 
multi-component chiral models,  
in particular in the presence of an isometry large enough, 
remains a future problem.
Our method will give a simple way to construct 
higher derivative non-linear sigma models 
on K\"ahler manifolds 
by gauging chiral fields with 
flat target spaces for which 
auxiliary field equations of motions are easy to solve.
In the strong gauge coupling limit, 
vector superfields do not have gauge kinetic terms 
becoming auxiliary superfields, and can be eliminated 
by their equations of motion.
This procedure is known as the K\"ahler quotients, 
see Ref.~\cite{Higashijima:1999ki} for constructions of 
hermitian symmetric spaces.
Thus, it will be possible to construct 
higher-derivative non-linear sigma models 
on hermitian symmetric spaces, 
as a generalization of the Faddeev-Skyrme 
${\mathbb C}P^1$ model. 

In this paper, we have not introduced superpotentials 
while we introduced them for non-gauged chiral models 
in our previous paper \cite{Nitta:2014pwa}. 
In the presence of a superpotential, there are more 
varieties of BPS topological solitons such as 
domain walls 
\cite{Dvali:1996xe}
in $U(N)$ gauge theories \cite{Isozumi:2004jc}, 
domain wall junctions  
\cite{GiTo,Oda:1999az} 
or networks \cite{Eto:2005cp}, and 
vortices ending on or stretched between domain walls 
\cite{Gauntlett:2000de,Isozumi:2004vg}.
In these cases, the auxiliary field equation can be solved 
at most perturbatively even for a single component, 
as was so for non-gauged chiral models \cite{Nitta:2014pwa}.

We also comment that in our gauged model, $\Lambda_{ik\bar{j}\bar{l}}$
does not contain space-time derivatives of the chiral superfields, 
unlike the non-gauged cases for which it is possible as for 
the supersymmetric Dirac-Born-Infeld action in 
Eq.~\eqref{eq:Lambda_DBI} and
the supersymmetric Faddeev-Skyrme model in Eq.~(\ref{eq:FSmodel}). 
A simple gauge covariant generalization of the form
\eqref{eq:Lambda_DBI} or (\ref{eq:FSmodel}) does not provide supersymmetric interactions of the vector superfield. 
It is interesting to introduce the gauge covariant derivatives of $\Phi$
in a supersymmetric way in the K\"ahler tensor $\Lambda_{ik\bar{j}\bar{l}}$,  
in order to construct a gauged Dirac-Born-Infeld action
\cite{Sasaki:2009ij} or a gauged Faddeev-Skyrme model.

In Ref.~\cite{Eto:2005sw},
1/2, 1/4, and 1/8 BPS states  
were classified in 
${\cal N}=2$ supersymmetric field theories 
without higher derivative terms.  
Extension to ${\cal N}=2$ supersymmetric field theories 
with higher derivative terms should be an interesting future problem.
In particular, 
1/4 BPS states in the canonical branch 
may have 1/8 BPS state counterparts   
in the non-canonical branch. 
While off-shell supersymmetry for eight supercharges 
is a hard task because one needs harmonic superfield 
or projective superfield formalisms,
partially off-shell supersymmetry that 
BPS solitons preserve can be used 
to construct an effective theory of BPS solitons \cite{Eto:2006uw}.

Extension to supergravity is also interesting 
for application to cosmology such as the ghost condensations 
and the Galileon inflation models in supersymmetric theories 
along the line in Refs.~\cite{KhLeOv}--\cite{FaKe}.

\subsection*{Acknowledgments}
The work of M.\ N.\ is supported in part by Grant-in-Aid for 
Scientific Research (No. 25400268) 
from the Ministry of Education, Culture, Sports, Science and Technology  (MEXT) of Japan.
The work of S.~S. is supported in part by Kitasato University Research Grant for Young
Researchers.

\begin{appendix}
\section{Notation and conventions}\label{sec:notation}
We use the convention in the textbook of 
Wess and Bagger \cite{Wess:1992cp}. 
The component expansion of the $\mathcal{N} = 1$ chiral superfield in
the $x$-basis is 
\begin{equation}
\Phi (x, \theta, \bar{\theta}) = \varphi 
+ i \theta \sigma^m \bar{\theta} \partial_m \varphi + \frac{1}{4}
\theta^2 \bar{\theta}^2 \Box \varphi + \theta^2 F,
\end{equation}
where 
only the bosonic components are presented.
The supercovariant derivatives are defined as 
\begin{eqnarray}
D_{\alpha} = \frac{\partial}{\partial \theta^{\alpha}} + i
 (\sigma^m)_{\alpha \dot{\alpha}} \bar{\theta}^{\dot{\alpha}}
 \partial_m, \quad 
\bar{D}_{\dot{\alpha}} = - \frac{\partial}{\partial
\bar{\theta}^{\dot{\alpha}}} - i \theta^{\alpha} (\sigma^m)_{\alpha
\dot{\alpha}} \partial_m.
\end{eqnarray}
The sigma matrices are $\sigma^m = (\mathbf{1}, \vec{\tau})$.
Here $\vec{\tau} = (\tau^1, \tau^2, \tau^3)$ are Pauli matrices.
The bosonic components of the supercovariant derivatives of $\Phi^i$ are 
\begin{align}
D^{\alpha} \Phi^i D_{\alpha} \Phi^j =& \ 
- 4 \bar{\theta}^2 \partial_m \varphi^i \partial^m \varphi^j 
+ 4 i (\theta \sigma^m \bar{\theta}) (\partial_m \varphi^i F^j + F^i
 \partial_m \varphi^j) 
- 4 \theta^2 F^i F^j 
\notag \\
& \ + 2 \theta^2 \bar{\theta}^2 
\left(
\Box \varphi^i F^j + F^i \Box \varphi^j - \partial_m \varphi^i
 \partial^m F^j - \partial_m F^i \partial^m \varphi^j
\right), \\
\bar{D}_{\dot{\alpha}} \Phi^{\dagger\bar{i}} \bar{D}^{\dot{\alpha}}
 \Phi^{\dagger\bar{j}} =& \ 
- 4 \theta^2 \partial_m \bar{\varphi}^{\bar{i}} \partial^m
 \bar{\varphi}^{\bar{j}} 
- 4 i (\theta \sigma^m \bar{\theta}) (\partial_m \bar{\varphi}^{\bar{i}}
 \bar{F}^{\bar{j}} + \bar{F}^{\bar{i}} \partial_m
 \bar{\varphi}^{\bar{j}}) 
+ 4 \bar{\theta}^2 \bar{F}^{\bar{i}} \bar{F}^{\bar{j}}
\notag \\
& \ 
+ 2 \theta^2 \bar{\theta}^2 
\left(
\bar{F}^{\bar{i}} \Box \bar{\varphi}^{\bar{j}} + \Box
 \bar{\varphi}^{\bar{i}} \bar{F}^{\bar{j}} 
- \partial_m \bar{\varphi}^{\bar{i}} \partial^m \bar{F}^{\bar{j}}
- \partial_m \bar{F}^{\bar{i}} \partial^m \bar{\varphi}^{\bar{j}}
\right), 
\\
D^{\alpha} \Phi^i D_{\alpha} \Phi^k \bar{D}_{\dot{\alpha}}
 \Phi^{\dagger\bar{j}} \bar{D}^{\dot{\alpha}}
 \Phi^{\dagger\bar{l}} 
=& \ 16 \theta^2 \bar{\theta}^2 
\left[
\frac{}{}
(\partial_m \varphi^i \partial^m \varphi^k) (\partial_m
 \bar{\varphi}^{\bar{j}} \partial^m \bar{\varphi}^{\bar{l}})
\right. 
\notag \\
& 
\left.
- \frac{1}{2} 
\left(
\partial_m \varphi^i F^k + F^i \partial_m \varphi^k 
\right)
\left(
\partial^n \bar{\varphi}^{\bar{j}} \bar{F}^{\bar{l}} 
+ \bar{F}^{\bar{j}} \partial^n \bar{\varphi}^{\bar{l}}
\right)
+ F^i \bar{F}^{\bar{j}} F^k \bar{F}^{\bar{l}}
\right].
\end{align}
When the supercovariant derivative is gauged, we obtain
\begin{align}
\mathcal{D}_{\alpha} \Phi =& \ 
2 i (\sigma^m)_{\alpha \dot{\alpha}} \bar{\theta}^{\dot{\alpha}} D_m
 \varphi + 2 \theta_{\alpha} F + 2 \theta_{\alpha} \bar{\theta}^2 (\Box
 \varphi + g D \varphi) - \frac{1}{2} (\sigma^m)_{\alpha \dot{\alpha}}
 (\bar{\sigma}^n)^{\dot{\alpha}} {}_{\dot{\beta}} \theta^{\beta}
 \bar{\theta}^2 (\partial_m \partial_n \varphi 
\notag \\
& \ 
- 2 i  g \partial_m A_n
 \varphi) + i \theta^2 (\sigma^m)_{\alpha \dot{\alpha}}
 \bar{\theta}^{\dot{\alpha}} \partial_m F.
\end{align}
Using this expression, we obtain Eq.~\eqref{eq:4th_deri_gauge2_component}.

\section{Supersymmetry variation of fermions}\label{sec:SUSY_variation}
The explicit supersymmetry variation of the fermions in the Euclidean
space is given by 
\begin{align}
\delta_{\xi} \psi_{\alpha}^i =& \ \sqrt{2}i 
\left(
\begin{array}{c}
(\partial_4 + i \partial_3) \varphi^i \bar{\xi}^{\dot{1}} + i
 (\partial_1 - i \partial_2) \varphi^i \bar{\xi}^{\dot{2}} - i \xi_1
 F^i \\
(\partial_4 - i \partial_3) \varphi^i \bar{\xi}^{\dot{2}} + i
 (\partial_1 + i \partial_2) \varphi^i \bar{\xi}^{\dot{1}} - i \xi_2 F^i
\end{array}
\right), \\
\delta_{\xi} \bar{\psi}^{\dot{\alpha}i} =& \  \sqrt{2} i 
\left(
\begin{array}{c}
(\partial_4 - i \partial_3) \bar{\varphi}^i \xi_1 - i (\partial_1 - i
 \partial_2) \bar{\varphi}^i \xi_2 - i \bar{\xi}^{\dot{1}} \bar{F}^i \\
(\partial_4 + i \partial_3) \bar{\varphi}^i \xi_2 - i (\partial_1 + i
 \partial_2) \bar{\varphi}^i \xi_1 - i \bar{\xi}^{\dot{2}} \bar{F}^i
\end{array}
\right).
\end{align}
\begin{align}
\delta_{\xi} \lambda_{\alpha} =& \ 
\left(
\begin{array}{c}
i \xi_1 D + i \xi_1 (F_{12} + F_{34})
- \xi_2 (F_{13} - i F_{14} - i F_{23} - F_{24}) \\
i \xi_2 D - i \xi_2 (F_{12} + F_{34}) 
+  \xi_1 (F_{13} + i F_{14} + i F_{23} - F_{24})
\end{array}
\right), \\
\delta_{\xi} \bar{\lambda}^{\dot{\alpha}} =& \ 
\left(
\begin{array}{c}
- i \bar{\xi}^{\dot{1}} D - i \bar{\xi}^{\dot{1}} (F_{12} -
 F_{34}) + \bar{\xi}^{\dot{2}} (F_{13} + i F_{14} - i F_{23}
 +  F_{24} ) \\
- i \bar{\xi}^{\dot{2}} D + i \bar{\xi}^{\dot{2}} (F_{12} -
 F_{34}) - \bar{\xi}^{\dot{1}} (F_{13} - i F_{14} + i
 F_{23} + F_{24})
\end{array}
\right).
\end{align}

\end{appendix}



\begin{thebibliography}{0}
\bibitem{Leutwyler:1993iq} 
  H.~Leutwyler,
  ``On the foundations of chiral perturbation theory,''
  Annals Phys.\  {\bf 235}, 165 (1994)
  [hep-ph/9311274].

\bibitem{Seiberg:1994rs} 
  N.~Seiberg and E.~Witten,
  ``Electric - magnetic duality, monopole condensation, and confinement in N=2 supersymmetric Yang-Mills theory,''
  Nucl.\ Phys.\ B {\bf 426}, 19 (1994)
  [Erratum-ibid.\ B {\bf 430}, 485 (1994)]
  [hep-th/9407087],
  ``Monopoles, duality and chiral symmetry breaking in N=2 supersymmetric QCD,''
  Nucl.\ Phys.\ B {\bf 431}, 484 (1994)
  [hep-th/9408099].

\bibitem{Witten:1978mh} 
  E.~Witten and D.~I.~Olive,
  ``Supersymmetry Algebras That Include Topological Charges,''
  Phys.\ Lett.\ B {\bf 78}, 97 (1978).

\bibitem{Nitta:2014pwa} 
  M.~Nitta and S.~Sasaki,
  ``BPS States in Supersymmetric Chiral Models with Higher Derivative Terms,''
  Phys.\ Rev.\ D {\bf 90}, no. 10, 105001 (2014)
  [arXiv:1406.7647 [hep-th]].




\bibitem{Nemeschansky:1984cd} 
  D.~Nemeschansky and R.~Rohm,
  ``Anomaly Constraints On Supersymmetric Effective Lagrangians,''
  Nucl.\ Phys.\ B {\bf 249}, 157 (1985).

\bibitem{Gates:1995fx} 
  S.~J.~Gates, Jr.,
  ``Why auxiliary fields matter: The Strange case of the 4-D, N=1 supersymmetric QCD effective action,''
  Phys.\ Lett.\ B {\bf 365}, 132 (1996)
  [hep-th/9508153], 
  ``Why auxiliary fields matter: The strange case of the 4-D, N=1 supersymmetric QCD effective action. 2.,''
  Nucl.\ Phys.\ B {\bf 485}, 145 (1997)
  [hep-th/9606109].


\bibitem{Gates:2000rp} 
  S.~J.~Gates, Jr., M.~T.~Grisaru, M.~E.~Knutt and S.~Penati,
  ``The Superspace WZNW action for 4-D, N=1 supersymmetric QCD,''
  Phys.\ Lett.\ B {\bf 503}, 349 (2001)
  [hep-ph/0012301];
  S.~J.~Gates, Jr., M.~T.~Grisaru, M.~E.~Knutt, S.~Penati and H.~Suzuki,
  ``Supersymmetric gauge anomaly with general homotopic paths,''
  Nucl.\ Phys.\ B {\bf 596}, 315 (2001)
  [hep-th/0009192];
  S.~J.~Gates, Jr., M.~T.~Grisaru and S.~Penati,
  ``Holomorphy, minimal homotopy and the 4-D, N=1 supersymmetric Bardeen-Gross-Jackiw anomaly,''
  Phys.\ Lett.\ B {\bf 481}, 397 (2000)
  [hep-th/0002045].

\bibitem{Nitta:2001rh} 
  M.~Nitta,
  ``A Note on supersymmetric WZW term in four dimensions,''
  Mod.\ Phys.\ Lett.\ A {\bf 15}, 2327 (2000)
  [hep-th/0101166].


\bibitem{Buchbinder:1994iw}
  I.~L.~Buchbinder, S.~Kuzenko and Z.~Yarevskaya,
  ``Supersymmetric effective potential: Superfield approach,''
  Nucl.\ Phys.\ B {\bf 411}, 665 (1994).

\bibitem{Buchbinder:1994xq}
  I.~L.~Buchbinder, S.~M.~Kuzenko and A.~Y.~Petrov,
  ``Superfield chiral effective potential,''
  Phys.\ Lett.\ B {\bf 321} (1994) 372.

\bibitem{Matone:1996bj}
  M.~Matone,
  ``Modular invariance and structure of the exact Wilsonian action of N = 2 supersymmetric Yang-mills theory,''
  Phys.\ Rev.\ Lett.\  {\bf 78} (1997) 1412
  [hep-th/9610204].

\bibitem{Bellisai:1997ck}
  D.~Bellisai, F.~Fucito, M.~Matone and G.~Travaglini,
  ``Nonholomorphic terms in N=2 SUSY Wilsonian actions and RG equation,''
  Phys.\ Rev.\ D {\bf 56} (1997) 5218
  [hep-th/9706099].

\bibitem{Banin:2006db} 
  A.~T.~Banin, I.~L.~Buchbinder and N.~G.~Pletnev,
  ``On quantum properties of the four-dimensional generic chiral superfield model,''
  Phys.\ Rev.\ D {\bf 74}, 045010 (2006)
  [hep-th/0606242].

\bibitem{AnDuGh}
  I.~Antoniadis, E.~Dudas and D.~M.~Ghilencea,
  ``Supersymmetric Models with Higher Dimensional Operators,''
  JHEP {\bf 0803} (2008) 045
  [arXiv:0708.0383 [hep-th]].

\bibitem{Gomes:2009ev}
  M.~Gomes, J.~R.~Nascimento, A.~Y.~Petrov and A.~J.~da Silva,
  ``On the effective potential in higher-derivative superfield theories,''
  Phys.\ Lett.\ B {\bf 682} (2009) 229
  [arXiv:0908.0900 [hep-th]].

\bibitem{Gama:2011ws}
  F.~S.~Gama, M.~Gomes, J.~R.~Nascimento, A.~Y.~Petrov and A.~J.~da Silva,
  ``On the higher-derivative supersymmetric gauge theory,''
  Phys.\ Rev.\ D {\bf 84} (2011) 045001
  [arXiv:1101.0724 [hep-th]].

\bibitem{Kuzenko:2014ypa}
  S.~M.~Kuzenko and S.~J.~Tyler,
  ``The one-loop effective potential of the Wess-Zumino model revisited,''
  JHEP {\bf 1409} (2014) 135
  [arXiv:1407.5270 [hep-th]].

\bibitem{BeNeSc}
  E.~A.~Bergshoeff, R.~I.~Nepomechie and H.~J.~Schnitzer,
  ``Supersymmetric Skyrmions in Four-dimensions,''
  Nucl.\ Phys.\ B {\bf 249} (1985) 93.

\bibitem{Fr}
  L.~Freyhult,
  ``The Supersymmetric extension of the Faddeev model,''
  Nucl.\ Phys.\ B {\bf 681} (2004) 65
  [hep-th/0310261].

\bibitem{RoTs}
  M.~Rocek and A.~A.~Tseytlin,
  ``Partial breaking of global D = 4 supersymmetry, constrained superfields, and three-brane actions,''
  Phys.\ Rev.\ D {\bf 59} (1999) 106001
  [hep-th/9811232].

\bibitem{SaYaYo}
  S.~Sasaki, M.~Yamaguchi and D.~Yokoyama,
  ``Supersymmetric DBI inflation,''
  Phys.\ Lett.\ B {\bf 718} (2012) 1
  [arXiv:1205.1353 [hep-th]].


\bibitem{AdQuSaGuWe3}
  C.~Adam, J.~M.~Queiruga, J.~Sanchez-Guillen and A.~Wereszczynski,
  ``Supersymmetric K field theories and defect structures,''
  Phys.\ Rev.\ D {\bf 84} (2011) 065032
  [arXiv:1107.4370 [hep-th]].

\bibitem{AdQuSaGuWe2}
  C.~Adam, J.~M.~Queiruga, J.~Sanchez-Guillen and A.~Wereszczynski,
  ``BPS bounds in supersymmetric extensions of K field theories,''
  Phys.\ Rev.\ D {\bf 86} (2012) 105009
  [arXiv:1209.6060 [hep-th]].

\bibitem{Eto:2012qda} 
  M.~Eto, T.~Fujimori, M.~Nitta, K.~Ohashi and N.~Sakai,
  ``Higher Derivative Corrections to Non-Abelian Vortex Effective Theory,''
  Prog.\ Theor.\ Phys.\  {\bf 128}, 67 (2012)
  [arXiv:1204.0773 [hep-th]].

\bibitem{Adam:2011hj} 
  C.~Adam, J.~M.~Queiruga, J.~Sanchez-Guillen and A.~Wereszczynski,
  ``N=1 supersymmetric extension of the baby Skyrme model,''
  Phys.\ Rev.\ D {\bf 84}, 025008 (2011)
  [arXiv:1105.1168 [hep-th]].

\bibitem{AdQuSaGuWe}
  C.~Adam, J.~M.~Queiruga, J.~Sanchez-Guillen and A.~Wereszczynski,
  ``Extended Supersymmetry and BPS solutions in baby Skyrme models,''
  JHEP {\bf 1305} (2013) 108
  [arXiv:1304.0774 [hep-th]].

\bibitem{Bolognesi:2014ova} 
  S.~Bolognesi and W.~Zakrzewski,
  ``Baby Skyrme Model, Near-BPS Approximations and Supersymmetric Extensions,''
  Phys.\ Rev.\ D {\bf 91}, no. 4, 045034 (2015)
  [arXiv:1407.3140 [hep-th]].

\bibitem{Nitta:2014fca}
 M.~Nitta and S.~Sasaki,
 ``Higher Derivative Corrections to Manifestly Supersymmetric
Nonlinear Realizations,''
 Phys.\ Rev.\ D {\bf 90}, 105002 (2014)
 [arXiv:1408.4210 [hep-th]].




\bibitem{KhLeOv}
  J.~Khoury, J.~-L.~Lehners and B.~Ovrut,
  ``Supersymmetric P(X,$\phi$) and the Ghost Condensate,''
  Phys.\ Rev.\ D {\bf 83} (2011) 125031
  [arXiv:1012.3748 [hep-th]].

\bibitem{KoLeOv2}
  M.~Koehn, J.~-L.~Lehners and B.~Ovrut,
  ``Ghost condensate in $N=1$ supergravity,''
  Phys.\ Rev.\ D {\bf 87} (2013) 6,  065022
  [arXiv:1212.2185 [hep-th]].

\bibitem{KhLeOv2}
  J.~Khoury, J.~-L.~Lehners and B.~A.~Ovrut,
  ``Supersymmetric Galileons,''
  Phys.\ Rev.\ D {\bf 84} (2011) 043521
  [arXiv:1103.0003 [hep-th]].

\bibitem{KoLeOv}
  M.~Koehn, J.~-L.~Lehners and B.~A.~Ovrut,
  ``Higher-Derivative Chiral Superfield Actions Coupled to N=1 Supergravity,''
  Phys.\ Rev.\ D {\bf 86} (2012) 085019
  [arXiv:1207.3798 [hep-th]].

\bibitem{FaKe}
  F.~Farakos and A.~Kehagias,
  ``Emerging Potentials in Higher-Derivative Gauged Chiral Models Coupled to N=1 Supergravity,''
  JHEP {\bf 1211} (2012) 077
  [arXiv:1207.4767 [hep-th]].


\bibitem{Adam:2009px} 
  C.~Adam, P.~Klimas, J.~Sanchez-Guillen and A.~Wereszczynski,
  ``Compact baby skyrmions,''
  Phys.\ Rev.\ D {\bf 80}, 105013 (2009)
  [arXiv:0909.2505 [hep-th]];
  C.~Adam, T.~Romanczukiewicz, J.~Sanchez-Guillen and A.~Wereszczynski,
  ``Investigation of restricted baby Skyrme models,''
  Phys.\ Rev.\ D {\bf 81} (2010) 085007
  [arXiv:1002.0851 [hep-th]];
  C.~Adam, J.~Sanchez-Guillen, A.~Wereszczynski and W.~J.~Zakrzewski,
  ``Topological duality between vortices and planar Skyrmions in BPS theories with area-preserving diffeomorphism symmetries,''
  Phys.\ Rev.\ D {\bf 87}, no. 2, 027703 (2013)
  [arXiv:1209.5403 [hep-th]];
  C.~Adam, T.~Romanczukiewicz, J.~Sanchez-Guillen and A.~Wereszczynski,
  ``Magnetothermodynamics of BPS baby skyrmions,''
  JHEP {\bf 1411}, 095 (2014)
  [arXiv:1405.5215 [hep-th]].


\bibitem{Adam:2012pm} 
  C.~Adam, C.~Naya, J.~Sanchez-Guillen and A.~Wereszczynski,
  ``The gauged BPS baby Skyrme model,''
  Phys.\ Rev.\ D {\bf 86}, 045010 (2012)
  [arXiv:1205.1532 [hep-th]];
  C.~Adam, C.~Naya, T.~Romanczukiewicz, J.~Sanchez-Guillen and A.~Wereszczynski,
  ``Topological phase transitions in the gauged BPS baby Skyrme model,''
  arXiv:1501.03817 [hep-th].




\bibitem{Abrikosov:1956sx} 
  A.~A.~Abrikosov,
  ``On the Magnetic properties of superconductors of the second group,''
  Sov.\ Phys.\ JETP {\bf 5}, 1174 (1957)
  [Zh.\ Eksp.\ Teor.\ Fiz.\  {\bf 32}, 1442 (1957)];
  H.~B.~Nielsen and P.~Olesen,
  ``Vortex Line Models for Dual Strings,''
  Nucl.\ Phys.\ B {\bf 61}, 45 (1973).


\bibitem{Hanany:2003hp} 
  A.~Hanany and D.~Tong,
  ``Vortices, instantons and branes,''
  JHEP {\bf 0307}, 037 (2003)
  [hep-th/0306150];
  R.~Auzzi, S.~Bolognesi, J.~Evslin, K.~Konishi and A.~Yung,
  ``NonAbelian superconductors: Vortices and confinement in N=2 SQCD,''
  Nucl.\ Phys.\ B {\bf 673}, 187 (2003)
  [hep-th/0307287];
  M.~Eto, Y.~Isozumi, M.~Nitta, K.~Ohashi and N.~Sakai,
  ``Moduli space of non-Abelian vortices,''
  Phys.\ Rev.\ Lett.\  {\bf 96}, 161601 (2006)
  [hep-th/0511088];
  M.~Eto, K.~Konishi, G.~Marmorini, M.~Nitta, K.~Ohashi, W.~Vinci and N.~Yokoi,
  ``Non-Abelian Vortices of Higher Winding Numbers,''
  Phys.\ Rev.\ D {\bf 74}, 065021 (2006)
  [hep-th/0607070].

\bibitem{Polyakov:1975yp} 
  A.~M.~Polyakov and A.~A.~Belavin,
  ``Metastable States of Two-Dimensional Isotropic Ferromagnets,''
  JETP Lett.\  {\bf 22}, 245 (1975)
  [Pisma Zh.\ Eksp.\ Teor.\ Fiz.\  {\bf 22}, 503 (1975)].

\bibitem{Schroers:1995he} 
  B.~J.~Schroers,
  ``Bogomolny solitons in a gauged O(3) sigma model,''
  Phys.\ Lett.\ B {\bf 356}, 291 (1995)
  [hep-th/9506004];
  B.~J.~Schroers,
  ``The Spectrum of Bogomol'nyi solitons in gauged linear sigma models,''
  Nucl.\ Phys.\ B {\bf 475}, 440 (1996)
  [hep-th/9603101];
  J.~M.~Baptista,
  ``Vortex equations in Abelian gauged sigma-models,''
  Commun.\ Math.\ Phys.\  {\bf 261}, 161 (2006)
  [math/0411517 [math-dg]];
  A.~Alonso-Izquierdo, W.~G.~Fuertes and J.~M.~Guilarte,
  ``Two species of vortices in massive gauged non-linear sigma models,''
  JHEP {\bf 1502}, 139 (2015)
  [arXiv:1409.8419 [hep-th]].

\bibitem{Nitta:2011um} 
  M.~Nitta and W.~Vinci,
  ``Decomposing Instantons in Two Dimensions,''
  J.\ Phys.\ A {\bf 45}, 175401 (2012)
  [arXiv:1108.5742 [hep-th]].

\bibitem{Brecher:1998tv}
  D.~Brecher,
  ``BPS states of the nonAbelian Born-Infeld action,''
  Phys.\ Lett.\ B {\bf 442} (1998) 117
  [hep-th/9804180].

\bibitem{Adam:2008rf} 
  C.~Adam, P.~Klimas, J.~Sanchez-Guillen and A.~Wereszczynski,
  ``Compact gauge K vortices,''
  J.\ Phys.\ A {\bf 42}, 135401 (2009)
  [arXiv:0811.4503 [hep-th]].

\bibitem{Ostrogradski}
  M.~Ostrogradski, ``Memoires sur les equations differentielles relatives au probleme des
  isoperimetres,'' 
  Mem. \ Ac. \ St. Petersbourg VI (1850) 385.

\bibitem{Hanany:2004ea} 
  A.~Hanany and D.~Tong,
  ``Vortex strings and four-dimensional gauge dynamics,''
  JHEP {\bf 0404}, 066 (2004)
  [hep-th/0403158].


\bibitem{Eto:2004rz} 
  M.~Eto, Y.~Isozumi, M.~Nitta, K.~Ohashi and N.~Sakai,
  ``Instantons in the Higgs phase,''
  Phys.\ Rev.\ D {\bf 72}, 025011 (2005)
  [hep-th/0412048].

\bibitem{Eto:2006pg} 
  M.~Eto, Y.~Isozumi, M.~Nitta, K.~Ohashi and N.~Sakai,
  ``Solitons in the Higgs phase: The Moduli matrix approach,''
  J.\ Phys.\ A {\bf 39}, R315 (2006)
  [hep-th/0602170].


\bibitem{Fujimori:2008ee} 
  T.~Fujimori, M.~Nitta, K.~Ohta, N.~Sakai and M.~Yamazaki,
  ``Intersecting Solitons, Amoeba and Tropical Geometry,''
  Phys.\ Rev.\ D {\bf 78}, 105004 (2008)
  [arXiv:0805.1194 [hep-th]].



\bibitem{Wess:1992cp} 
  J.~Wess and J.~Bagger,
  ``Supersymmetry and supergravity,''
  Princeton, USA: Univ. Pr. (1992) 259 p

\bibitem{Higashijima:1999ki} 
  K.~Higashijima and M.~Nitta,
  ``Supersymmetric nonlinear sigma models as gauge theories,''
  Prog.\ Theor.\ Phys.\  {\bf 103}, 635 (2000)
  [hep-th/9911139].


\bibitem{Dvali:1996xe} 
  G.~R.~Dvali and M.~A.~Shifman,
  ``Domain walls in strongly coupled theories,''
  Phys.\ Lett.\ B {\bf 396}, 64 (1997)
  [Erratum-ibid.\ B {\bf 407}, 452 (1997)]
  [hep-th/9612128].

\bibitem{Isozumi:2004jc} 
  Y.~Isozumi, M.~Nitta, K.~Ohashi and N.~Sakai,
  ``Construction of non-Abelian walls and their complete moduli space,''
  Phys.\ Rev.\ Lett.\  {\bf 93}, 161601 (2004)
  [hep-th/0404198];
  Y.~Isozumi, M.~Nitta, K.~Ohashi and N.~Sakai,
  ``Non-Abelian walls in supersymmetric gauge theories,''
  Phys.\ Rev.\ D {\bf 70}, 125014 (2004)
  [hep-th/0405194];
  M.~Eto, Y.~Isozumi, M.~Nitta, K.~Ohashi, K.~Ohta and N.~Sakai,
  ``D-brane construction for non-Abelian walls,''
  Phys.\ Rev.\ D {\bf 71}, 125006 (2005)
  [hep-th/0412024];
  M.~Eto, Y.~Isozumi, M.~Nitta, K.~Ohashi, K.~Ohta, N.~Sakai and Y.~Tachikawa,
  ``Global structure of moduli space for BPS walls,''
  Phys.\ Rev.\ D {\bf 71}, 105009 (2005)
  [hep-th/0503033].
\bibitem{GiTo}
  G.~W.~Gibbons and P.~K.~Townsend,
  ``A Bogomolny equation for intersecting domain walls,''
  Phys.\ Rev.\ Lett.\  {\bf 83} (1999) 1727
  [hep-th/9905196];
  S.~M.~Carroll, S.~Hellerman and M.~Trodden,
  ``Domain wall junctions are 1/4 - BPS states,''
  Phys.\ Rev.\ D {\bf 61} (2000) 065001
  [hep-th/9905217];
  A.~Gorsky and M.~A.~Shifman,
  ``More on the tensorial central charges in N=1 supersymmetric gauge theories (BPS wall junctions and strings),''
  Phys.\ Rev.\ D {\bf 61}, 085001 (2000)
  [hep-th/9909015].

\bibitem{Oda:1999az} 
  H.~Oda, K.~Ito, M.~Naganuma and N.~Sakai,
  ``An Exact solution of BPS domain wall junction,''
  Phys.\ Lett.\ B {\bf 471}, 140 (1999)
  [hep-th/9910095]; 
  K.~Ito, M.~Naganuma, H.~Oda and N.~Sakai,
  ``Nonnormalizable zero modes on BPS junctions,''
  Nucl.\ Phys.\ B {\bf 586}, 231 (2000)
  [hep-th/0004188];
  M.~Naganuma, M.~Nitta and N.~Sakai,
  ``BPS walls and junctions in SUSY nonlinear sigma models,''
  Phys.\ Rev.\ D {\bf 65}, 045016 (2002)
  [hep-th/0108179].

\bibitem{Eto:2005cp} 
  M.~Eto, Y.~Isozumi, M.~Nitta, K.~Ohashi and N.~Sakai,
  ``Webs of walls,''
  Phys.\ Rev.\ D {\bf 72}, 085004 (2005)
  [hep-th/0506135],
  ``Non-Abelian webs of walls,''
  Phys.\ Lett.\ B {\bf 632}, 384 (2006)
  [hep-th/0508241];
  M.~Eto, T.~Fujimori, T.~Nagashima, M.~Nitta, K.~Ohashi and N.~Sakai,
  ``Dynamics of Domain Wall Networks,''
  Phys.\ Rev.\ D {\bf 76}, 125025 (2007)
  [arXiv:0707.3267 [hep-th]].


\bibitem{Gauntlett:2000de} 
  J.~P.~Gauntlett, R.~Portugues, D.~Tong and P.~K.~Townsend,
  ``D-brane solitons in supersymmetric sigma models,''
  Phys.\ Rev.\ D {\bf 63}, 085002 (2001)
  [hep-th/0008221]; 
  M.~Shifman and A.~Yung,
  ``Domain walls and flux tubes in N=2 SQCD: D-brane prototypes,''
  Phys.\ Rev.\ D {\bf 67}, 125007 (2003)
  [hep-th/0212293].

\bibitem{Isozumi:2004vg} 
  Y.~Isozumi, M.~Nitta, K.~Ohashi and N.~Sakai,
  ``All exact solutions of a 1/4 Bogomol'nyi-Prasad-Sommerfield equation,''
  Phys.\ Rev.\ D {\bf 71}, 065018 (2005)
  [hep-th/0405129];
  M.~Eto, T.~Fujimori, T.~Nagashima, M.~Nitta, K.~Ohashi and N.~Sakai,
  ``Dynamics of Strings between Walls,''
  Phys.\ Rev.\ D {\bf 79}, 045015 (2009)
  [arXiv:0810.3495 [hep-th]].

\bibitem{Sasaki:2009ij}
  S.~Sasaki,
   ``On Non-linear Action for Gauged M2-brane,''
   JHEP {\bf 1002} (2010) 039
   [arXiv:0912.0903 [hep-th]].

\bibitem{Eto:2005sw} 
  M.~Eto, Y.~Isozumi, M.~Nitta and K.~Ohashi,
  ``1/2, 1/4 and 1/8 BPS equations in SUSY Yang-Mills-Higgs systems: Field theoretical brane configurations,''
  Nucl.\ Phys.\ B {\bf 752}, 140 (2006)
  [hep-th/0506257].

\bibitem{Eto:2006uw} 
  M.~Eto, Y.~Isozumi, M.~Nitta, K.~Ohashi and N.~Sakai,
  ``Manifestly supersymmetric effective Lagrangians on BPS solitons,''
  Phys.\ Rev.\ D {\bf 73}, 125008 (2006)
  [hep-th/0602289].

\end{thebibliography}
\end{document}